\documentclass[a4paper,11pt]{article}
\pdfoutput=1
\usepackage[T1]{fontenc}

\usepackage{jheppub}

\usepackage{cancel}
\usepackage{mathrsfs}
\usepackage{amsmath}   
\usepackage{amssymb}   
\usepackage{feynmp-auto,expdlist}
\usepackage{float}
\usepackage[style=numeric-comp,sorting=none,subentry, natbib=true]{biblatex}
 \newcommand{\beq}[1]{\begin{equation}\label{#1}}
 \newcommand{\eeq}{\end{equation}}
 \newcommand{\bea}[1]{\begin{eqnarray}\label{#1}}
 \newcommand{\eea}{\end{eqnarray}}

\title{Torsion induced one-loop corrections to inflaton decay and the Stochastic gravitational waves}
 \author[a]{AlexKen Lee}
 \author[a]{Keyun Wu}
 \affiliation[a]{Departament de F\'{i}sica Qu\`{a}ntica i Astrof\'{i}sica, Institut de Ci\`{e}ncies del Cosmos (ICCUB), Universitat de Barcelona, Mart\'{i} i Franqu\`{e}s 1, E-08028 Barcelona, Spain}
\emailAdd{alexkenlee@163.com}
\emailAdd{keyunwu@fqa.ub.edu}

\abstract{We investigate one-loop corrections from torsion-induced four-fermion interactions to inflaton three-body decay and their impact on the associated stochastic gravitational-wave signal. We find a pronounced asymmetry in the dependence on the renormalization scale $u$. While the enhancement of the gravitational-wave spectrum remains modest, not exceeding roughly a factor of order unity for representative inflaton masses well below the Planck scale within the perturbative regime, the suppression can be much stronger, reaching up to two orders of magnitude, corresponding to reductions at the percent level. These results imply that loop corrections, particularly fermionic self-interactions, can significantly reduce the predicted gravitational-wave signal in models based on tree-level analyses. This suppression may shift the signal outside the sensitivity range of future observations and should therefore be taken into account in realistic phenomenological studies.}

\begin{document} 
\maketitle
\flushbottom

\section{Introduction}

When fermions are coupled to gravity, their intrinsic spin requires the vierbein $e_\mu^{~a}$ and the spin connection $\omega_{\mu ab}$ to be treated as independent dynamical variables. This framework is commonly referred to as the first-order formalism of gravity with fermions, or equivalently the Einstein–Cartan–Sciama–Kibble theory \cite{Cartan:1923zea,Cartan:1924yea,Kibble:1961ba,Hehl:1976kj,Hehl:1974cn,Freedman:2012zz}. In this formulation, the equation of motion for the spin connection naturally gives rise to spacetime torsion. Physically, the effects of torsion can be reinterpreted as inducing local four-fermion self-interactions. From an effective field theory (EFT) perspective, such interactions correspond to dimension-six operators suppressed by the Planck scale. The interplay between fermions and torsion in curved spacetime has motivated extensive investigations and has led to a variety of intriguing implications in cosmology and astrophysics. These include modifications to gravitational collapse and potential mechanisms for singularity avoidance \cite{Kerlick:1975tr,Ziaie:2013pma,Bambi:2014uua}, scenarios for non-singular cosmological evolution \cite{Gasperini:1986mv,Kopczynski:1972fhu,Gasperini:1998eb,Magueijo:2012ug,Alexander:2014eva,Cabral:2020mzw}, fermion condensation driven by torsion-induced interactions during reheating, reminiscent of the Bardeen–Cooper–Schrieffer mechanism \cite{Alexander:2008vt,Weller:2013iga,Tong:2023krn}, interpretations of torsion as a candidate for dark matter or dark energy \cite{Alexander:2009uu,Poplawski:2011wj,Belyaev:2016icc}, possible insights into the cosmological constant problem \cite{Alexander:2005vb,Alexander:2006we}, and the generation of quantum entanglement in expanding spacetime with torsion-induced corrections \cite{Belfiglio:2021mnr}. Despite the rich phenomenology predicted by torsion and its associated four-fermion EFT interactions, direct observational signatures of these effects remain scarce to date. 

Motivated by recent studies of stochastic gravitational-wave (GW) signals originating from inflaton decay \cite{Nakayama:2018ptw,Huang:2019lgd,Barman:2023ymn,Barman:2023rpg,Kanemura:2023pnv,Hu:2024awd,Xu:2024fjl,Inui:2024wgj,Jiang:2024akb,Lee:2025lyk,Lee:2026fpg,Tokareva:2023mrt,Choi:2024ilx,Kaneta:2024yyn,Mantziris:2024uzz,Datta:2024tne,Bernal:2024jim,Xu:2025wjq,Banik:2025olw,Ai:2025fqw,Wang:2025lmf,Blasi:2024vew}, our previous work investigated gravitational-wave production from many-body inflaton decay processes mediated by torsion-induced four-fermion interactions \cite{Lee:2025lyk}. As a natural extension of Ref.~\cite{Lee:2025lyk}, the present work focuses on the role of such interactions in inflaton three-body decay channels. This necessarily requires the inclusion of radiative corrections, and in this paper we restrict our analysis to the leading one-loop contributions. It is worth emphasizing that loop effects in this context have rarely been explored in the existing literature. The main reason is that, when the inflaton mass lies well below the Planck scale, loop-induced contributions from higher-dimensional operators are typically expected to be strongly suppressed and thus phenomenologically negligible, particularly regarding any enhancement of the gravitational-wave spectrum. For instance, the one-loop induced inflaton two-body decay into a pair of gravitons within the framework of Yukawa interactions and minimally coupled gravity, $\varphi \to h_{ab}h_{mn}$, was studied in Ref.~\cite{Xu:2024fjl}. Notably, this process arises purely at the loop level, since minimal gravitational coupling does not provide a tree-level vertex for double-graviton emission from inflaton decay. It was shown that this loop-induced two-body channel is significantly subdominant compared to the tree-level three-body (bremsstrahlung) decay and the tree-level $2\to2$ scattering processes. On the other hand, when the inflaton mass approaches or exceeds the Planck scale, quantum-gravitational effects are expected to become important. However, in this regime, the absence of a complete and systematically controlled theory of quantum gravity limits the reliability of perturbative predictions. The parameter region explored in the present work therefore represents a useful intermediate window: it allows for a controlled analysis within an effective field theory framework, while still capturing potentially non-negligible quantum effects induced by torsion. From this perspective, it is particularly instructive to consider the regime in which the inflaton mass, $M$, lies approximately one order of magnitude below the Planck scale, $M \sim 0.1\,M_{\rm Pl}$, where $M_{\rm Pl} \simeq 2.4 \times 10^{18}\,\mathrm{GeV}$ denotes the reduced Planck mass. This choice provides a tractable setting to probe the qualitative impact of torsion-induced loop corrections on observable quantities such as the gravitational-wave spectrum. Moreover, our analysis is partially inspired by the well-known Coleman–Weinberg mechanism, in which the running of the renormalization scale $u$ plays a central role in shaping the effective potential \cite{Coleman:1973jx}. Indeed, for physically reasonable choices of the running scale, a variety of nontrivial phenomena in inflationary model building have been uncovered \cite{Drees:2021wgd,Drees:2022aea,Ballesteros:2015noa,Germani:2014hqa,DiVita:2015bha,Fumagalli:2020ody,Fumagalli:2016sof}. In a similar spirit, the present work explores how the running of $u$, induced by torsion-generated loop corrections, can affect inflaton decay processes and their associated stochastic gravitational-wave signatures.

Building on the above analysis and motivations, the structure of this work is organized as follows. In Sec.~\ref{OneLoopTwoBodyDecayRate}, we first provide a brief overview of the torsion-induced four-fermion interaction and analyze the corresponding one-loop corrections to the two-body inflaton decay process with a fermion–antifermion pair in the final state. In particular, we introduce a dimensionless quantity $R^{(0)}_{\varphi\to\bar{\psi}\psi}$, defined as the ratio between the squared amplitude including one-loop corrections and its tree-level counterpart. This quantity serves as a diagnostic for quantifying the relative size of loop effects and allows us to constrain the viable range of the renormalization scale $u$. Subsequently, in Sec.~\ref{OneLoopThreeBodyDecayRate}, we extend this analysis to the inflaton three-body decay channel. In close analogy, we introduce a corresponding dimensionless ratio $R^{(1)}_{\varphi\to\bar{\psi}\psi h_{ab}}$ to characterize the impact of loop corrections in this process. Since the stochastic gravitational-wave spectrum is directly related to the quantity $\chi=\frac{d\Gamma_{\varphi\to\bar{\psi}\psi h_{ab}}^{(1)}/dE_{l}}{\Gamma_{\varphi\to\bar{\psi}\psi}^{(0)}}$, our focus is to investigate how the running of the renormalization scale $u$ affects $\chi$, while restricting to parameter regions in which both $R^{(0)}{\varphi\to\bar{\psi}\psi}$ and $R^{(1)}{\varphi\to\bar{\psi}\psi h}$ remain within a controlled and perturbatively reliable regime. By systematically identifying the maximal deviation of $\chi_{\text{tree+one-loop}}$ relative to its tree-level counterpart $\chi_{\text{tree}}$ under the running of $u$, we then proceed in Sec.~\ref{StochasticGWs} to determine the corresponding maximal deviation induced in the stochastic gravitational-wave spectrum. Finally, in Sec.~\ref{ConAndDiscuss}, we present further physical interpretations and discuss possible future directions. Additional technical details, including the evaluation of scattering amplitudes and fermion loop integrals, are provided in Appendices~\ref{IdentityToFermiLoopIntegral}–\ref{ThreeBodyAmpliOneLoop}.

\section{Setup of the theoretical model and one-loop corrected two-body decay \label{OneLoopTwoBodyDecayRate}}

Let us now take into account both the torsion–induced four–fermion interaction and the Yukawa interaction, namely
\begin{align}
\nonumber
S_{\text{total}}&=\int d^{4}x\,\big\{\partial^{a}\varphi\partial_{a}\varphi-M^{2}\varphi^{2}+\text{i}\bar{\psi}\gamma^{a}\partial_{a}\psi-m_{\psi}\bar{\psi}\psi \\
&-y_\psi\varphi\bar{\psi}\psi+\underbrace{\frac{\kappa^{2}}{64}\bar{\psi}\gamma_{mnl}\psi\times\bar{\psi}\gamma^{mnl}\psi}_{\mathcal{L}_{\text{torsion-}4f}}\big\},
\end{align}
where $y_{\psi}$ denotes a dimensionless coupling constant, and $\kappa \equiv \sqrt{16\pi}/M_{\rm Pl}$ is the gravitational coupling.

\begin{figure}[ht]
	\begin{center}
	\includegraphics[width=0.845\columnwidth]{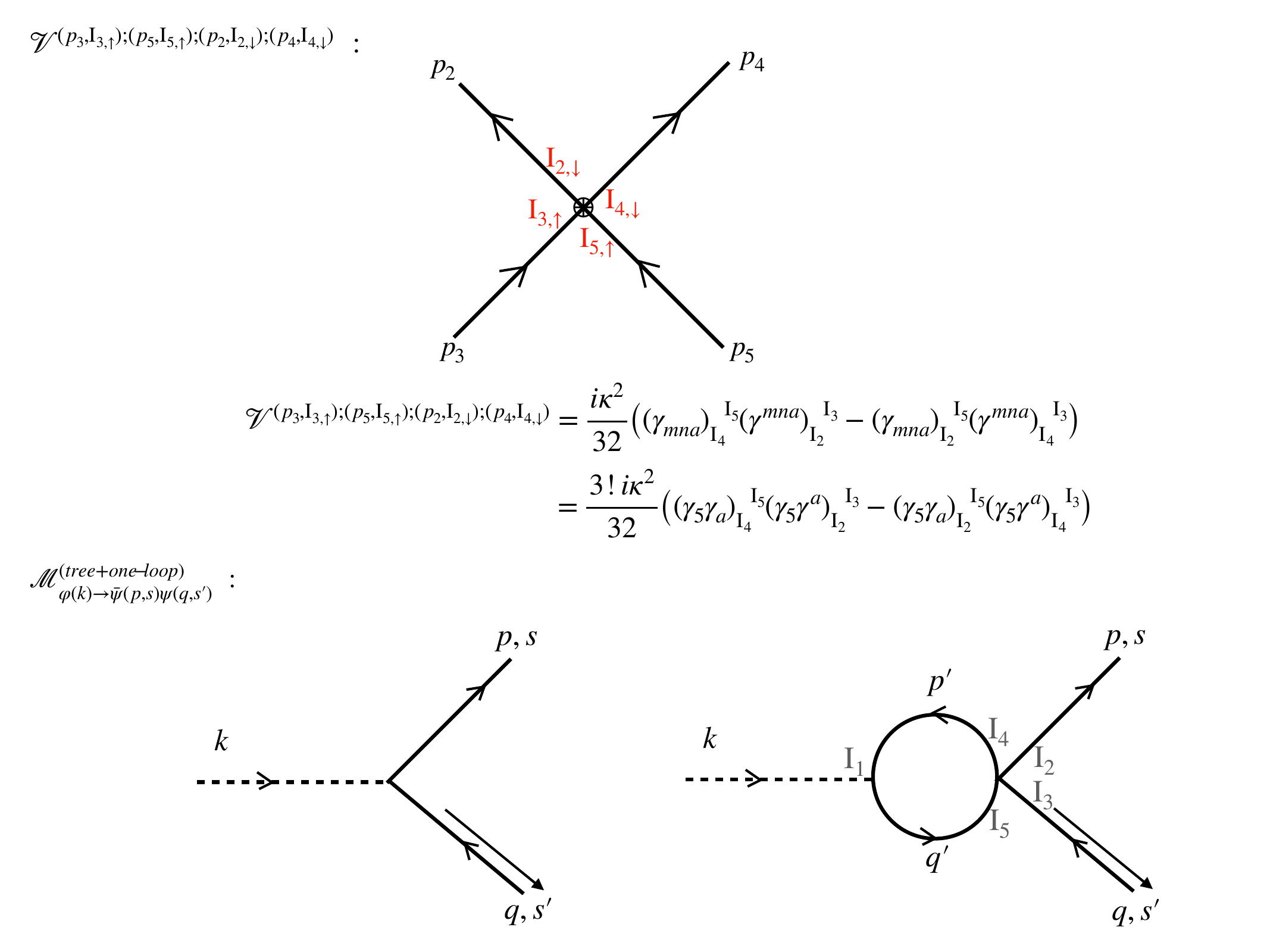}
		\caption{The upper panel shows the Feynman rules associated with the torsion-induced four-fermion interaction, while the lower panel illustrates the one-loop Feynman diagrams for the two-body decay process mediated jointly by the Yukawa interaction and the torsion-induced four-fermion interaction.}
		\label{TorsionOneLoopTwoBody}
	\end{center}
\end{figure}

According to the dual relations of the complete Clifford algebra in four–dimensional spacetime \cite{Freedman:2012zz,Lee:2025lyk}, the torsion–induced four–fermion interaction can also be rewritten in the equivalent form
\begin{align}
&\mathcal{L}_{\text{torsion-}4f}=\frac{3!\kappa^{2}}{64}\bar{\psi}(\gamma_{5}\gamma_{a})\psi\times\bar{\psi}(\gamma_{5}\gamma^{a})\psi.
\end{align}The Feynman rule associated with the interaction term $\mathcal{L}_{\text{torsion-}4f}$ is illustrated in Fig.~\ref{TorsionOneLoopTwoBody}. In particular, a detailed derivation of the torsion–induced four–fermion interaction within the first–order (Palatini) formalism for gravity coupled to fermions, as well as a systematic discussion of the Feynman rules represented in Fig.~\ref{TorsionOneLoopTwoBody}, can be found in our previous works \cite{Lee:2025lyk}. Here, the four-momentum $k$ denotes the on-shell momentum of the inflaton, while $p$ and $q$ correspond to the four-momenta of the external fermions in the final state. The quantities $p'$ and $q'$ represent the off-shell loop momenta.  The amplitude for the two-body decay induced jointly by the Yukawa interaction and the torsion-induced four-fermion interaction (as depicted in the lower panel of Fig.~\ref{TorsionOneLoopTwoBody}) with the one-loop corrections can be written as Eq. \eqref{AmplitudeTwoBodyOneLoopv1}. As shown in Appendix \ref{IdentityToFermiLoopIntegral}, the trace contribution in the first term of the amplitude Eq. Eq. \eqref{AmplitudeTwoBodyOneLoopv1} vanishes. Therefore, we only need to deal with the second term in the amplitude Eq. \eqref{AmplitudeTwoBodyOneLoopv1}. After evaluating the loop integral using dimensional regularization and the minimal subtraction scheme ($\overline{\text{MS}}$), we obtain the expression Eq. \eqref{AmplitudeTwoBodyOneLoopv2}, where $u$ denotes the renormalization energy scale.

\begin{align}
\nonumber
\mathcal{M}_{\varphi(k)\to\bar{\psi}(p,s)\psi(q,s^{\prime})}^{\text{(tree+one-loop)}}&=y_{\psi}\bar{u}(\boldsymbol{p},s)v(\boldsymbol{q},s^{\prime})\bigg(\frac{3\kappa^{2}}{16}\{(2M^{2}-8m_{\psi}^{2})F_{1}(M^{2};m_{\psi},m_{\psi})\\
\label{AmplitudeTwoBodyOneLoopv2}
&+2(M^{2}-6m_{\psi}^{2})\log(\frac{u^{2}}{m_{\psi}^{2}})+3M^{2}-14m_{\psi}^{2}\}-\text{i}\bigg),
\end{align}
in which $M$ and $m_{\psi}$ denote the masses of the inflaton and the final-state fermions, respectively.  Note that the function $F_1(s;a,b)$ takes the general form
\begin{align}
&\Delta =s^2+(a^2-b^2)^2-2s(a^2+b^2)~,~F_1(s;a,b)=\frac{\left(\Delta\right)^{1/2}}{s}\tanh^{-1}\left(\frac{\Delta^{1/2}}{a^2+b^2-s}\right).
\end{align}Accordingly, the specific case $F_1(M^2;m_{\psi},m_{\psi})$ appearing in Eq. \eqref{AmplitudeTwoBodyOneLoopv2} admits the explicit expression
\begin{align}
\label{DefineF1}
&F_1(M^2;m_{\psi},m_{\psi})=(1-4m_{\psi}^2/M^2)^{1/2}\,\text{arctanh}\left(\frac{M^2(1-4m_{\psi}^2/M^2)^{1/2}}{2m_{\psi}^2-M^2}\right).
\end{align}With the result in Eq. \eqref{AmplitudeTwoBodyOneLoopv2}, the corresponding squared amplitude for the decay process $\varphi \to \psi\bar{\psi}$, incorporating both the tree-level contribution and the one-loop corrections, is given by
\begin{align}
\nonumber
\sum_{s}\sum_{s^\prime}\vert\mathcal{M}_{\varphi(k)\to\bar{\psi}(p,s)\psi(q,s^{\prime})}^{\text{(tree+one-loop)}}\vert^{2}&=\frac{9\kappa^4y_{\psi}^{2}}{128}\big(M^2-4m_\psi^{2}\big)\\ \nonumber
&\hspace{-2cm}\times\bigg(-2m_\psi^{2}\big(4F_{1}(M^{2};m_{\psi},m_{\psi})+6\log(\frac{u^{2}}{m_{\psi}^{2}})+7\big)\\
\label{SquareAmpliTwoBodyTreeLoop}
&\hspace{-2cm}+M^{2}\big(2F_{1}(M^{2};m_{\psi},m_{\psi})+2\log(\frac{u^{2}}{m_{\psi}^{2}})+3\big)\bigg)^{2}+2y_{\psi}^2(M^2-4m^2_\psi).
\end{align}From Eq. \eqref{SquareAmpliTwoBodyTreeLoop}, the corresponding two-body decay rate can be obtained
\begin{align}
\Gamma_{\varphi\to\bar{\psi}\psi}^{(0)}=\frac{ y_{\psi}^2 M \left(1-4y^2\right)^{3/2}}{2048\pi}\times\left(256+9M^4(B_1-2A_{1}y^2)^2\kappa^4\right),
\end{align}where the shorthand functions $A_{1}(M,m_{\psi},u)$ and $B_{1}(M,m_{\psi},u)$ are introduced to represent 
\begin{align}
&A_{1}(M,m_{\psi},u)=4F_{1}(M^{2};m_{\psi},m_{\psi})+6\log(\frac{u^{2}}{m_{\psi}^{2}})+7,\\
&B_{1}(M,m_{\psi},u)=2F_{1}(M^{2};m_{\psi},m_{\psi})+2\log(\frac{u^{2}}{m_{\psi}^{2}})+3.
\end{align}
However, in this part our primary interest concentrates on the dimensionless ratio between the decay rate including one-loop corrections and the decay rate computed at tree level only, namely
\begin{align}
\nonumber
R^{(0)}_{\varphi\to\bar{\psi}\psi}&=\frac{\sum_{s}\sum_{s^\prime}\vert\mathcal{M}_{\varphi(k)\to\bar{\psi}(p,s)\psi(q,s^{\prime})}^{\text{(tree+one-loop)}}\vert^{2}}{\sum_{s}\sum_{s^\prime}\vert\mathcal{M}_{\varphi(k)\to\bar{\psi}(p,s)\psi(q,s^{\prime})}^{\text{(tree)}}\vert^{2}}-1\\
\label{TwoBodyDecayPerturJudge}
&\hspace{-0.6cm}=9\pi^{2}\lambda^{4}\bigg(3-14y^{2}+2(1-4y^{2})^{3/2}\text{arccoth}(\frac{2y^{2}-1}{\sqrt{1-4y^{2}}})+(4-24y^{2})\text{log}(x)\bigg)^{2}.
\end{align}

\begin{figure}[ht]
 	\begin{center}
 		\includegraphics[scale=0.369]{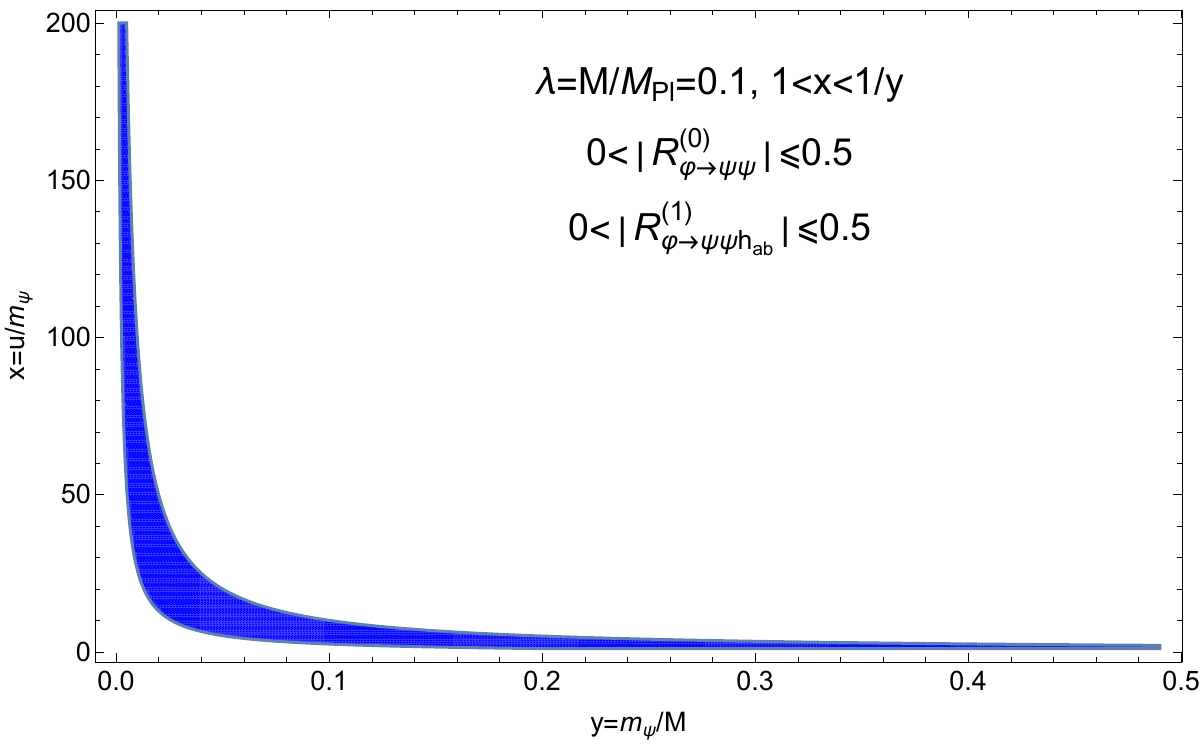}
        \includegraphics[scale=0.369]{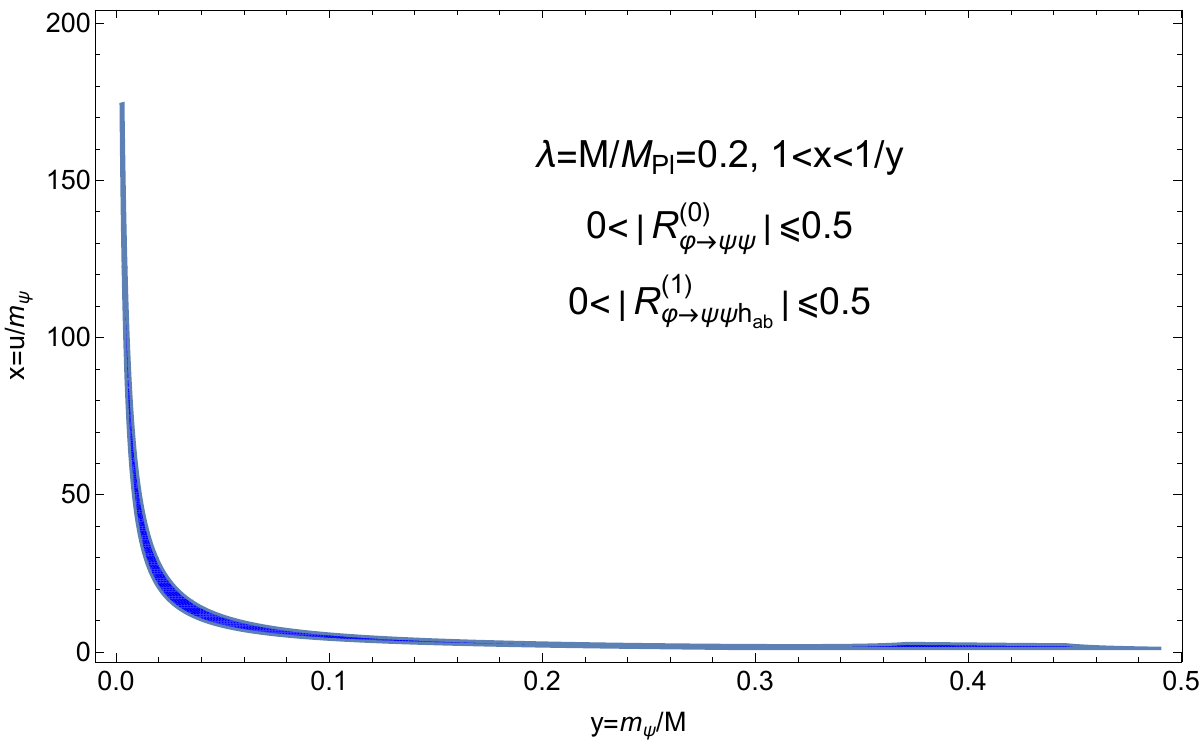}
 		\caption{Combined constraints on the allowed parameter space in the $x=u/m_\psi$ and $y=m_\psi/M$ plane, derived from the ratios $R^{(0)}_{\varphi\to\bar{\psi}\psi}$ in~Eq. \eqref{TwoBodyDecayPerturJudge} for the inflaton two-body decay and $R_{\varphi\to\bar{\psi}\psi h_{ab}}^{(1)}$ in~Eq. \eqref{DifferThreeBodyDecayPerturJudge} for the three-body decay channel. The imposed criterion requires that the squared amplitude, including both tree-level and one-loop contributions, remains within the range $[0.5,1.5]$ of the tree-level result.}
 		\label{TwoBodyAndThreeBodyRenorScale}
 	\end{center}
 \end{figure}

In the above, we have introduced three dimensionless parameters $y,\lambda,x$, which characterize the ratios among the relevant energy scales, namely $y = m_\psi / M$, $\lambda = M / M_{\rm Pl}$, and $x = u / m_\psi$. To be specific, the purpose of introducing the quantity Eq. \eqref{TwoBodyDecayPerturJudge} is to employ the constraints $\vert R^{(0)}_{\varphi\to\bar{\psi}\psi}(\lambda,y,x)\vert<1$ and $1<x<1/y$ to approximately delineate a viable region of parameter space that maintains the perturbative validity of the QFT framework. The latter condition follows from the requirement that the renormalization scale $u$ should be greater than or equal to the mass of the heaviest particle in the system and less than or equal to that of the lightest particle, which implies $m_\psi \leq u \leq M$. In principle, the perturbativity bound of the loop expansion should be assessed by more rigorous methods, such as partial-wave perturbative unitarity bounds \cite{DuasoPueyo:2024usw,Goldberg:1990qk}. However, in the present work we take a more pragmatic approach: we assume that the perturbative QFT framework remains valid at least up to the one-loop level, and we use this assumption to delineate a reasonable range for the renormalization scale $u$, once some of the parameters among $m_\psi$, $M$, and $M_{\rm Pl}$ are fixed. Moreover, in the next section, when analyzing the differential three-body decay, we will compute a dimensionless ratio analogous to Eq. \eqref{TwoBodyDecayPerturJudge}, namely Eq. \eqref{DifferThreeBodyDecayPerturJudge}. Based on the expressions Eq. \eqref{TwoBodyDecayPerturJudge} and Eq. \eqref{DifferThreeBodyDecayPerturJudge}, we present in Fig.~\ref{TwoBodyAndThreeBodyRenorScale} the viable parameter region in the $x$–$y$ plane for a fixed value of $M$, obtained by imposing the requirement that perturbativity holds at least up to the one-loop level for both the two-body and three-body decay channels. With this admissible parameter space at hand, in Sec. \ref{StochasticGWs}, we will subsequently investigate the associated stochastic gravitational-wave signals by varying $u$ within this allowed domain, and systematically examine how the choice of renormalization scale influences the resulting gravitational-wave spectrum as well as the corresponding characteristic frequencies.

\begin{figure}[ht]
	\begin{center}
		\includegraphics[scale=0.43]{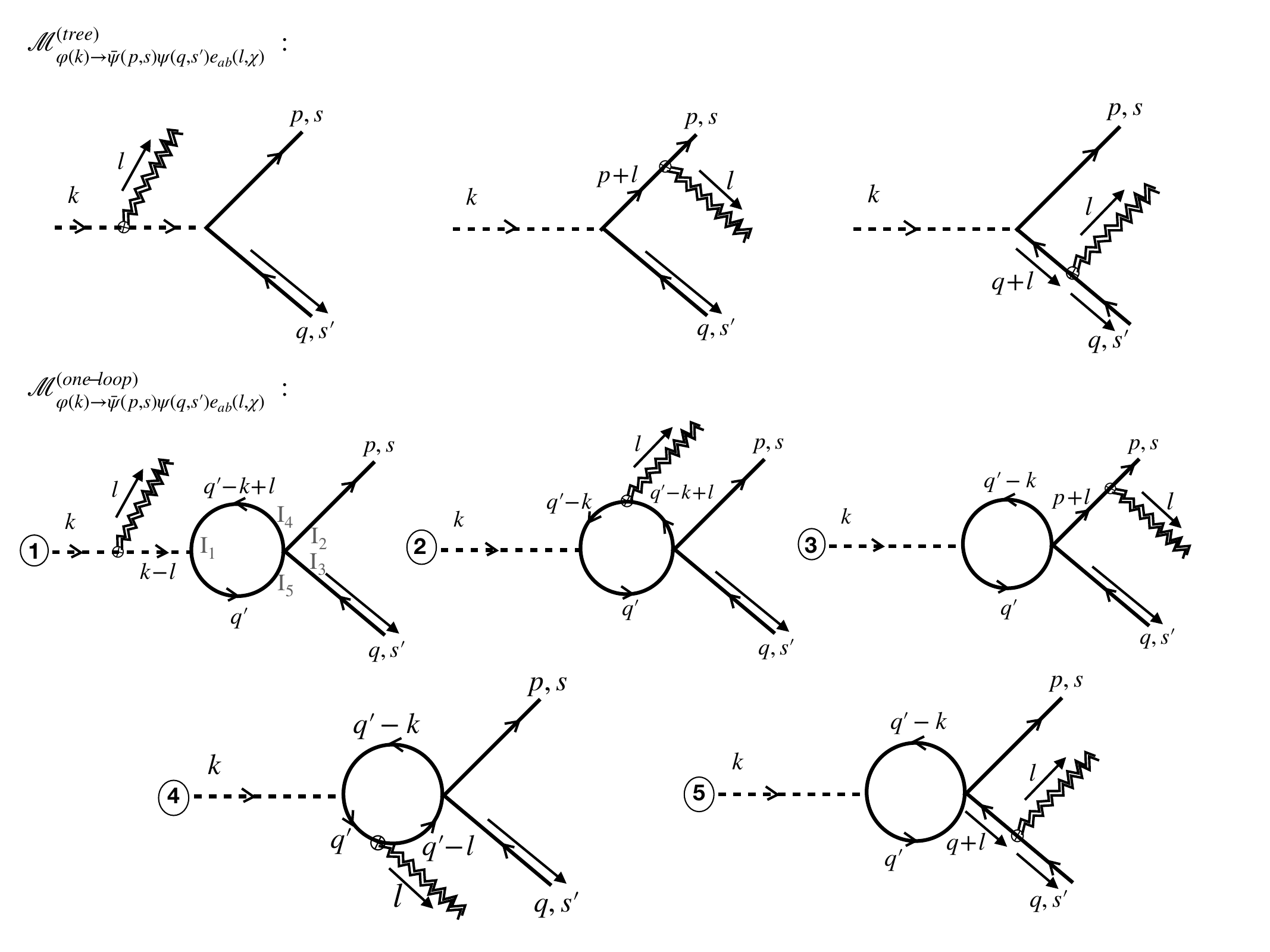}
		\caption{Feynman diagrams corresponding to the three-body decay of the inflaton, including fermionic one-loop corrections arising from torsion-induced four-fermion interactions, accompanied by the emission of a single graviton.}
		\label{TorsionOneLoopThreeBodyGWs}
	\end{center}
\end{figure}

\section{Differential three-body decay through a fermionic loop \label{OneLoopThreeBodyDecayRate}}

Subsequently, we proceed to incorporate the loop corrections to the three-body decay process involving the emission of a single graviton. Before proceeding further, we first present the amplitude Eq. \eqref{ThreeBodyDecayTree} corresponding to the tree-level inflaton decay process \cite{Nakayama:2018ptw,Bernal:2023wus}. We then consider the Feynman diagrams associated with the loop corrections to the three-body decay process arising from the torsion-induced four-fermion interaction, which are shown in Fig. \ref{TorsionOneLoopThreeBodyGWs}, with the corresponding amplitudes given in Eqs. \eqref{ThreeBodyLoopAmplitude1}--\eqref{ThreeBodyLoopAmplitude5}. Here, $l$ denotes the four-momentum of the final-state graviton. Note that the above results have been considerably simplified through extensive use of identities involving the $\gamma$ matrices; several useful identities are collected in Appendix~\ref{IdentityToFermiLoopIntegral}. In addition, one can visit \cite{Freedman:2012zz,Lee:2025lyk} for further details on the complete Clifford algebra in four-dimensional spacetime. Note that the five terms in Eqs. Eq. \eqref{ThreeBodyLoopAmplitude1}--Eq. \eqref{ThreeBodyLoopAmplitude5}, written from top to bottom, correspond one-to-one, in the same order, to the labeled Feynman diagrams in Fig. \ref{TorsionOneLoopThreeBodyGWs}. For each term, the result obtained after evaluating the loop integrals using dimensional regularization and the minimal subtraction scheme ($\overline{\text{MS}}$) is presented explicitly in Appendix~\ref{ThreeBodyAmpliOneLoop}. Here, we present only the final result for the modulus squared of the total three-body decay amplitude, which incorporates both the tree-level and one-loop contributions, as given in Eq. \eqref{SquareAmpliThreeBodyTreeLoop}. It is straightforward to verify that the effects of the loop corrections are entirely encoded in the overall factor appearing in the first line of Eq. \eqref{SquareAmpliThreeBodyTreeLoop}. Importantly, this factor is independent of the energy of the emitted graviton, which is physically consistent with the underlying structure of the interaction. At the one-loop level, the amplitudes corresponding to diagrams 1, 2, and 4 in Fig.~\ref{TorsionOneLoopThreeBodyGWs} vanish identically. Similarly, at tree level, the first diagram in Fig.~\ref{TorsionOneLoopThreeBodyGWs} also yields a zero contribution. Consequently, the squared amplitude in Eq. \eqref{SquareAmpliThreeBodyTreeLoop} effectively receives contributions only from diagrams 3 and 5 in Fig.~\ref{TorsionOneLoopThreeBodyGWs} at the loop level, together with the second and third diagrams at tree level. It is worth emphasizing that all these non-vanishing contributions share the same structural form at the amplitude level. This common structure explains why the loop effects can be factorized into a single overall multiplicative factor, as displayed in the first line of Eq. \eqref{SquareAmpliThreeBodyTreeLoop}. Furthermore, it is important to stress that, for the three-body decay process, graviton emission does not interfere with the loop integration. As a result, after applying dimensional regularization and the minimal subtraction scheme, the energy scale associated with the emitted graviton, $E_l$, does not enter the loop-induced contributions. For this reason, we assume that the inflaton two-body decay with one-loop corrections (corresponding to Fig.~\ref{TorsionOneLoopTwoBody}) and the three-body decay with one-loop corrections (corresponding to Fig.~\ref{TorsionOneLoopThreeBodyGWs}) share the same renormalization scale $u$, with $m_\psi \leq u \leq M$. Finally, we recall that the function $F_1$ appearing in the result has been defined in Eq. \eqref{DefineF1}.
\begin{align}
\nonumber
\sum_{s}\sum_{s^{\prime}}&\sum_{\chi}\vert \!\mathcal{M}_{\varphi(k)\to\bar{\psi}(p,s)\psi(q,s^{\prime})h_{ab}(l,\chi)}^{\text{(tree+one-loop)}}\!\vert^{2}\!=\frac{\kappa^{2}y_{\psi}^{2}}{1024E_{l}^{2}M^{2}(M-2E_{p})^{2}\left(M-2(E_{p}+E_{l})\right)^{2}}\\ \nonumber
&\times\bigg(6\kappa^{2}m_{\psi}^{2}\big\{4F_{1}+6\text{log}(\frac{u^{2}}{m_{\psi}^{2}})+7\big\}-3\kappa^{2}M^{2}\big\{2F_{1}+2\text{log}(\frac{u^{2}}{m_{\psi}^{2}})+3\big\}+16\bigg)^{2}\\ \nonumber
&\times\bigg(4M^{2}(E_{p}^{2}+3E_{p}E_{l}+E_{l}^{2})-4M^{3}(E_{p}+E_{l})-8E_{p}E_{l}M(E_{p}+E_{l})+4E_{l}^{2}m_{\psi}^{2}+M^{4}\bigg)\\ \nonumber
&\times\bigg(2E_{l}^{2}\left(M^{2}(4E_{p}^{2}-8E_{p}M+3M^{2})+2m_{\psi}^{2}M(8E_{p}-3M)-8m_{\psi}^{4}\right)+4E_{l}^{3}M^{2}(2E_{p}-M)\\ 
\label{SquareAmpliThreeBodyTreeLoop}
&+M^{2}(M-2E_{p})^{2}(M^{2}-4m_{\psi}^{2})-4E_{l}M(2E_{p}^{2}-3E_{p}M+M^{2})(M^{2}-4m_{\psi}^{2})\bigg),
\end{align}
in which $E_l$ and $E_p$ denote the energies of the massless graviton and the final-state fermion, respectively. The kinematic range is given by $0 < E_l < M/2$.

From the squared amplitude in Eq. \eqref{SquareAmpliThreeBodyTreeLoop}, one may construct the ratio of the complete result, incorporating both tree-level and one-loop contributions, to the tree-level contribution alone, as defined in Eq. \eqref{DifferThreeBodyDecayPerturJudge}. This ratio provides a convenient quantitative measure of the relative size of radiative corrections in the three-body decay channel. Following the same strategy adopted for the two-body decay process in Eq. \eqref{TwoBodyDecayPerturJudge}, we introduce the dimensionless quantity in Eq. \eqref{DifferThreeBodyDecayPerturJudge} as a criterion for perturbative consistency. In practice, it serves to identify the regions in the $x$–$y$ parameter plane where loop effects remain sufficiently suppressed, ensuring the reliability of the perturbative expansion underlying the squared amplitude Eq. \eqref{SquareAmpliThreeBodyTreeLoop}. The dimensionless parameters $\lambda$, $x$, and $y$ entering Eq. \eqref{DifferThreeBodyDecayPerturJudge} are defined in the same manner as in Eq. \eqref{TwoBodyDecayPerturJudge}. To further constrain the parameter space, we combine the condition derived from Eq. \eqref{TwoBodyDecayPerturJudge} with that from Eq. \eqref{DifferThreeBodyDecayPerturJudge}, and impose the requirement that the absolute values of the one-loop contributions to both the two-body and three-body decay channels do not exceed one half of their respective tree-level results. Under this perturbativity criterion, the allowed region in the $x$–$y$ plane, for a fixed mass scale $M$, can be identified from Fig.~\ref{TwoBodyAndThreeBodyRenorScale}. \begin{align}
\nonumber
R_{\varphi\to\bar{\psi}\psi e_{ab}}^{(1)}&=\frac{\sum_{\chi}\sum_{s}\sum_{s^{\prime}}\vert\mathcal{M}_{\varphi(k)\to\bar{\psi}(p,s)\psi(q,s^{\prime})h_{ab}(l,\chi)}^{\text{(tree+one-loop)}}\vert^{2}}{\sum_{\chi}\sum_{s}\sum_{s^{\prime}}\vert\mathcal{M}_{\varphi(k)\to\bar{\psi}(p,s)\psi(q,s^{\prime})h_{ab}(l,\chi)}^{\text{(tree)}}\vert^{2}}-1\\
\label{DifferThreeBodyDecayPerturJudge}
&=\bigg\{ 1+3\pi\lambda^{2}(14y^{2}-3)-6\pi\lambda^{2}\big((1-4y^{2})^{3/2}\text{arccoth}(\frac{2y^{2}-1}{\sqrt{1-4y^{2}}})\\ \nonumber
&+2(1-6y^{2})\text{log}(x)\big)^{2} \bigg\}^2-1.
\end{align}

\begin{figure}[H]
 	\begin{center}
 		\includegraphics[scale=0.8]{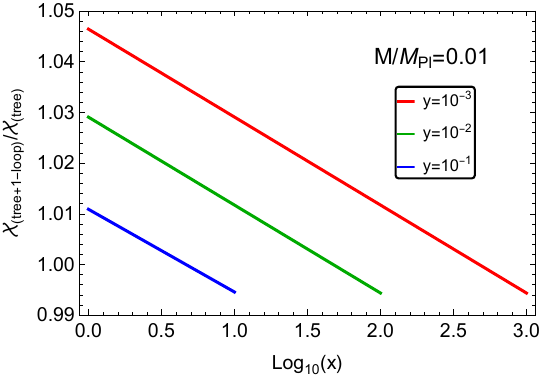}
        \includegraphics[scale=0.8]{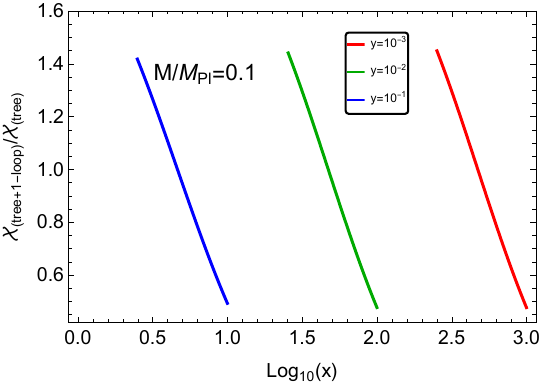}
 		\caption{This figure illustrates the deviation induced by the one-loop correction in
$\chi=\frac{d\Gamma_{\varphi\to\bar{\psi}\psi h_{ab}}^{(1)}/dE_{l}}{\Gamma_{\varphi\to\bar{\psi}\psi}^{(0)}}$
relative to its pure tree-level counterpart, evaluated within the parameter region in the $x$–$y$ plane allowed by Fig.~\ref{TwoBodyAndThreeBodyRenorScale} for different values of the inflaton mass $M$. Here we recall the definitions $x = u / m_\psi$ and $y = m_\psi / M$. In particular, the figure emphasizes how the magnitude of this deviation depends on the renormalization scale $u$ within the perturbatively admissible parameter space.}
 \label{MaxDeviationRenorScale}
 \end{center}
 \end{figure}

Using the squared amplitude in Eq. \eqref{SquareAmpliThreeBodyTreeLoop}, the corresponding differential three-body decay rate can be derived as Eq. \eqref{DifferThreeBodyWithLoop},
\begin{align}
\nonumber
\frac{d\Gamma_{\varphi\to\bar{\psi}\psi h_{ab}}^{(1)}}{dE_{l}}&=\frac{\kappa^{2}M^{2}y_{\psi}^{2}}{65536\pi^{3}}\\ \nonumber
&\times\bigg(3M^{2}\kappa^{2}\big\{2y^{2}(4F_{1}+12\log(x)+7)-(2F_{1}+4\log(x)+3)\big\}+16\bigg)^{2}\\
\label{DifferThreeBodyWithLoop}
&\times\!\!\bigg(\frac{(1-2z)\alpha}{z}\big(8zy^{2}+2z(z-1)-8y^{4}-2y^{2}+1\big)\\ \nonumber
&+\frac{4y^{2}\big((5-8z)y^{2}-(z-1)^{2}-4y^{4}\big)}{z}\ln(\frac{1+\alpha}{1-\alpha})\bigg).
\end{align}For notational convenience, we introduce the abbreviations $z = E_{l}/M$ and $\alpha=\sqrt{1-\frac{4y^{2}}{1-2z}}$. Since the gravitational-wave spectrum is proportional to the quantity
$\chi=\frac{d\Gamma_{\!\varphi\to\bar{\psi}\psi h_{ab}}^{(1)}\!/\!d\!E_{l}}{\Gamma_{\varphi\to\bar{\psi}\psi}^{(0)}}$,
our main interest is to examine how the inclusion of one-loop corrections modifies $\chi$ relative to its tree-level prediction within the parameter regions identified in Fig.~\ref{TwoBodyAndThreeBodyRenorScale}. More specifically, we aim to quantify the deviation of the ratio $\chi_{\text{tree+one-loop}} / \chi_{\text{tree}}$ from unity induced by the renormalization scale $u$. In practical calculations, the renormalization scale is conveniently parameterized by the dimensionless variable $x = u/m_{\psi}$. Our goal is therefore to determine the range of deviations of $\chi_{\text{tree+one-loop}} / \chi_{\text{tree}}$ from $1$ when $x$ lies within the region allowed by Fig.~\ref{TwoBodyAndThreeBodyRenorScale}. To this end, we first present the exact expression for the dimensionless ratio $\chi_{\text{tree+one-loop}}/\chi_{\text{tree}}$ given in Eq. \eqref{RatioRegardingChi},
\begin{align}
\nonumber
\chi_{\text{tree+one-loop}}\big / \chi_{\text{tree}}&=\big(\!\frac{d\Gamma_{\!\varphi\to\bar{\psi}\psi h_{ab}}^{(1)}\!/\!d\!E_{l}}{\Gamma_{\varphi\to\bar{\psi}\psi}^{(0)}}\!\big)_{\text{tree+one-loop}}\big/\!\big(\frac{d\Gamma_{\!\varphi\to\bar{\psi}\psi h_{ab}}^{(1)}\!/\! d\!E_{l}}{\Gamma_{\varphi\to\bar{\psi}\psi}^{(0)}}\big)_{\text{tree}}\\
\nonumber
&\hspace{-1cm}=\!1/256\big\{\!1\!+\!9\pi^{2}\lambda^{4}\big(3\!-\!14y^{2}\!+\!2(1\!-\!4y^{2})^{3/2}\!\text{arctanh}(\frac{\sqrt{1\!-\!4y^{2}}}{2y^{2}\!-\!1})\!+\!(4\!-\!24y^{2})\log x\big)^{2}\!\big\}\\
\label{RatioRegardingChi}
&\hspace{-1cm}\times\bigg(16\!+\!48\pi\lambda^{2}\big\{\!-\!3\!+\!2\sqrt{1\!-\!4y^{2}}\text{arctanh}(\frac{\sqrt{1-4y^{2}}}{1-2y^{2}})\!-\!4\log x\!\\ \nonumber
&\hspace{-1cm}+\!2y^{2}\big(7\!+\!4\sqrt{1\!-\!4y^{2}}\text{arctanh}(\frac{\sqrt{1-4y^{2}}}{2y^{2}-1})\!+\!12\log x\big)\big\}\bigg)^{2}.
\end{align} 
Based on this result, one can explicitly determine how this ratio varies with the renormalization scale $u$ while keeping the parameters $m_{\psi}$, $M_{\mathrm{Pl}}$, and $M$ fixed. The corresponding dependence is illustrated in Fig.~\ref{MaxDeviationRenorScale}. From Fig.~\ref{MaxDeviationRenorScale}, it is evident that when the inflaton mass satisfies $M/M_{\mathrm{Pl}} \lesssim 10^{-3}$, the loop-induced corrections remain numerically negligible and can be safely ignored for most practical purposes. In contrast, when the inflaton mass increases to the regime $M/M_{\mathrm{Pl}} \lesssim 10^{-1}$, noticeable deviations from the tree-level prediction begin to emerge. More specifically, for the case $M/M_{\mathrm{Pl}} \lesssim 10^{-1}$, we select several representative values of $y$ within the shaded region shown in the upper panel of Fig.~\ref{TwoBodyAndThreeBodyRenorScale}. Allowing the parameter $x$ to vary within the range permitted by this shaded region, one can then determine the behavior of $\chi_{\text{tree+one-loop}} / \chi_{\text{tree}}$ as a function of the renormalization scale $x$, while maintaining the validity of the perturbative expansion. It can be seen that, once the loop corrections are included, the quantity $\chi_{\text{tree+one-loop}}$ may deviate appreciably from its tree-level value. In particular, it can be enhanced to approximately $1.4 \,\chi_{\text{tree}}$. Moreover, it may also be suppressed to about $0.6 \, \chi_{\text{tree}}$, a feature that is somewhat unexpected. Motivated by the above analysis of how loop corrections and the renormalization scale affect the deviation of the decay rate from its tree-level prediction, in the next section we investigate in detail how the stochastic gravitational-wave signal generated during the inflaton decay process is influenced by the renormalization scale $u$ when torsion-induced loop corrections are taken into account.
 
\begin{figure}[H]
 	\begin{center}
 		\includegraphics[scale=0.363]{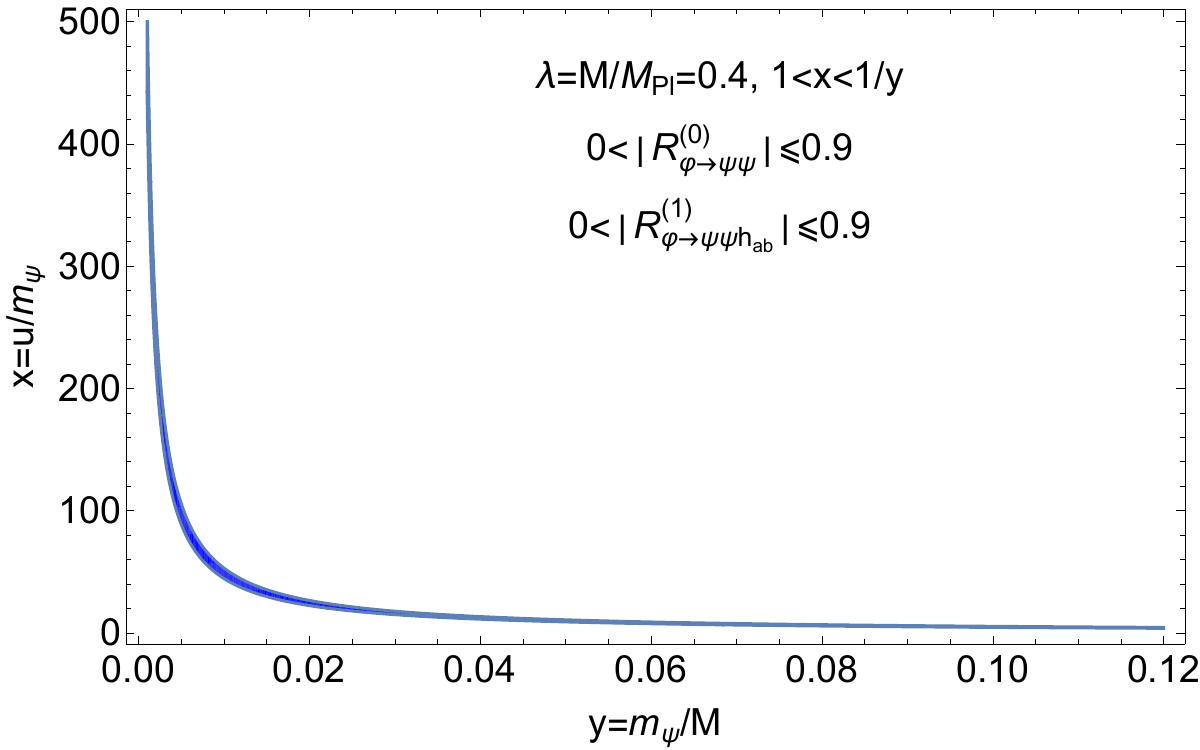}
        \includegraphics[scale=0.36]{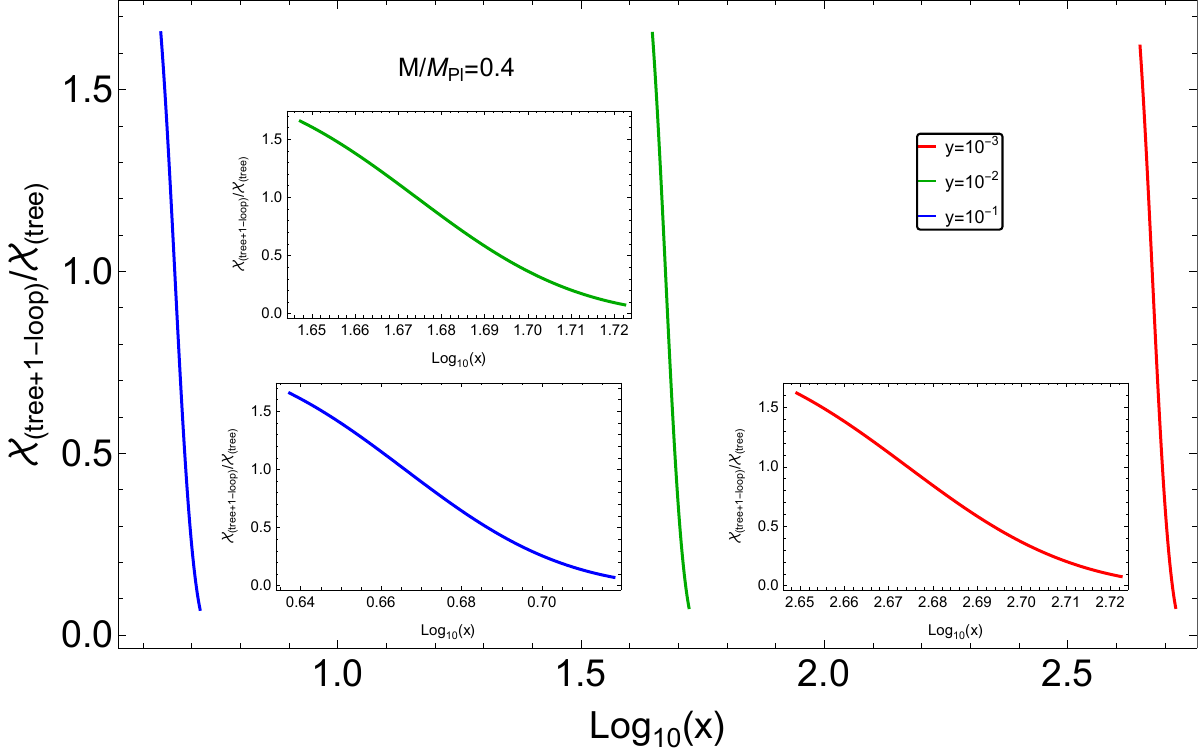}
 		\caption{Variation of the quantity
$\chi_{\text{tree+one-loop}}/\chi_{\text{tree}}$
as a function of the renormalization scale $u$, with the inflaton mass increased to $M=0.4M_{\text{pl}}$ and the perturbativity condition further relaxed.}
 \label{MaxDeviationRenorScaleRelaxPertur}
 \end{center}
 \end{figure}

 In addition, if the inflaton mass is moderately increased, for instance to $0.4M_{\rm Pl}$, and the perturbativity condition is further relaxed toward an extreme regime, as illustrated in the left panel of Fig.~\ref{MaxDeviationRenorScaleRelaxPertur}, one finds from the right panel of the same figure that, as the renormalization scale $u$ varies, the loop corrections enhance $\chi_{\text{tree+one-loop}}/\chi_{\text{tree}}$ by at most a factor of $\mathcal{O}(1.5)$. In contrast, the suppressive effect is considerably more pronounced: $\chi$ can be reduced by roughly half an order of magnitude and, in some cases, even by up to two orders of magnitude. We expect that this asymmetric impact of the running scale on $\chi$ will be reflected in the gravitational-wave spectrum, as will be discussed in Sec.~\ref{StochasticGWs}.

\section{Stochastic gravitional waves signals \label{StochasticGWs}}
Graviton emission can arise from bremsstrahlung in inflaton decays into fermionic final states. The total decay width can be decomposed as
\begin{align}
\Gamma_\varphi = \Gamma_\varphi^{(0)} + \Gamma_\varphi^{(1)},
\end{align}
where $\Gamma_\varphi^{(0)} \equiv \Gamma^{(0)}_{\varphi \to \bar{\psi}\psi}$ denotes the standard two-body decay width, while $\Gamma_\varphi^{(1)} \equiv \Gamma^{(1)}_{\varphi \to \bar{\psi}\psi l}$ corresponds to the three-body channel with graviton emission. Both contributions receive tree- and loop-level corrections.

The energy distribution of emitted gravitons, $d\Gamma_\varphi^{(1)}/dE_l$, enters the Boltzmann equations governing the evolution of energy densities,
\begin{align}
\dot{\rho}_\varphi + 3H\rho_\varphi &= -(\Gamma_\varphi^{(0)}+\Gamma_\varphi^{(1)})\rho_\varphi, \label{decayall}\\
\dot{\rho}_{\rm GW} + 4H\rho_{\rm GW} &= \int dE_l\, \frac{d\Gamma_\varphi^{(1)}}{dE_l}\frac{E_l}{M}\rho_\varphi, \\
\dot{\rho}_{\rm R} + 4H\rho_{\rm R} &= \Gamma_\varphi^{(0)}\rho_\varphi + \int dE_l\, \frac{d\Gamma_\varphi^{(1)}}{dE_l}\frac{M - E_l}{M}\rho_\varphi.
\end{align}
Here the factors $E_l/M$ and $(M-E_l)/M$ represent the fractions of inflaton energy carried by gravitational waves and radiation, respectively.
The total contribution from graviton-emitting processes can be written as
\begin{align}
\Gamma_\varphi^{(1)} \rho_\varphi 
= \int dE_l\, \frac{d\Gamma_\varphi^{(1)}}{dE_l}\rho_\varphi,
\end{align}
which can be decomposed into the energy injected into radiation and gravitational waves according to the above weights.
The differential rate develops an infrared divergence in the soft limit $E_l \to 0$~\cite{Lee:2025lyk}. To regulate this behavior, we introduce a lower cutoff $E_{l,\min} = 10^{-10}M$, while kinematics imposes an upper bound $E_l < M/2$. The integration range is therefore $10^{-10}M < E_l < M/2$.

We focus on the perturbative reheating regime, $a_{\rm Max} \ll a \ll a_{\rm rh}$ (equivalently $T_{\rm Max} \gg T \gg T_{\rm rh}$), in which the inflaton oscillates coherently and behaves effectively as pressureless matter. In this regime, the Hubble dilution dominates over decay, i.e.\ $3H\rho_\varphi \gg (\Gamma_\varphi^{(0)}+\Gamma_\varphi^{(1)})\rho_\varphi$, and Eq.~(\ref{decayall}) implies
\begin{align}
\rho_\varphi(a) \simeq \rho_\varphi(a_{\rm rh}) \left(\frac{a_{\rm rh}}{a}\right)^3.
\end{align}
The radiation temperature evolves as
\begin{align}
T(a) = T_{\rm rh}\left(\frac{a_{\rm rh}}{a}\right)^{3/8},
\end{align}
which leads to the scaling
\begin{align}
H(T) \simeq H(T_{\rm rh})\left(\frac{T}{T_{\rm rh}}\right)^4.
\end{align}

The evolution of the spectral ratio $d(\rho_{\rm GW}/\rho_{\rm R})/dE_l$ then follows
\begin{align}
\frac{d}{da}\frac{d(\rho_{\rm GW}/\rho_{\rm R})}{dE_l}
= \frac{1}{aH}\frac{\rho_\varphi}{\rho_{\rm R}}
\left[
\frac{d\Gamma_\varphi^{(1)}}{dE_l'}\frac{E_l'}{M}
- \Gamma_\varphi^{(0)} \frac{d(\rho_{\rm GW}/\rho_{\rm R})}{dE_l}
\right],
\end{align}
where $E_l'(a) = E_l (a_{\rm rh}/a)$ accounts for the redshift of the emitted gravitons.
Assuming $\Gamma_\varphi^{(1)} \ll \Gamma_\varphi^{(0)}$, reheating completes when $H(T_{\rm rh}) \simeq \Gamma_\varphi^{(0)}$. Taking vanishing initial radiation and GW densities, the solution yields
\begin{align}
\frac{d}{dE_l}\left(\frac{\rho_{\rm GW}}{\rho_{\rm R}}\right)_{T_{\rm rh}}
\simeq
\frac{1}{\Gamma_\varphi^{(0)}} \frac{d\Gamma_\varphi^{(1)}}{dE_l} \frac{E_l}{M}
\left[1 - \left(\frac{T_{\rm rh}}{T_{\rm Max}}\right)^{8/3}\right].
\end{align}
For $T_{\rm Max} \gg T_{\rm rh}$, the bracket approaches unity.

The present-day GW spectrum can be written as
\begin{align}
\Omega_{\rm GW} h^2
&= \Omega_\gamma^{(0)} h^2
\frac{g_*(T_{\rm rh})}{g_*(T_0)}
\left(\frac{g_{*s}(T_0)}{g_{*s}(T_{\rm rh})}\right)^{4/3}
\frac{1}{\Gamma_\varphi^{(0)}}
\frac{d\Gamma_\varphi^{(1)}}{dE_l}\frac{E_l}{M}
\left[1 - \left(\frac{T_{\rm rh}}{T_{\rm Max}}\right)^{8/3}\right] \, .
\end{align}
Using $d\ln f = d\ln E_l$, the graviton energy at reheating is related to the present-day frequency through
\begin{align}
E_l = 2\pi f \frac{T_{\rm rh}}{T_0}
\left(\frac{g_{*s}(T_{\rm rh})}{g_{*s}(T_0)}\right)^{1/3} \, .
\end{align}
Here $\Omega_\gamma^{(0)} h^2 \simeq 2.47 \times 10^{-5}$ denotes the current photon density parameter~\cite{Planck:2018vyg}, while $T_0 = 2.73\,\mathrm{K}$ is the present CMB temperature. The functions $g_*(T)$ and $g_{*s}(T)$ represent the effective relativistic degrees of freedom associated with energy density and entropy, respectively~\cite{Drees:2015exa}.

For the kinematic range $10^{-10}M < E_l < 0.5M$, and adopting $g_{*s}(T_0)=3.94$ and $g_{*s}(T_{\rm rh})=106.75$, the corresponding frequency range is bounded by
\begin{align}
f \lesssim 4.1 \times 10^{12}
\left(\frac{M}{M_{\rm Pl}}\right)
\left(\frac{5.5 \times 10^{15}\,\mathrm{GeV}}{T_{\rm rh}}\right)
\, \mathrm{Hz} \, .
\end{align}
This explicitly shows that both $M$ and $T_{\rm rh}$ set the observable GW frequency window.
The GW spectrum is governed by the dimensionless quantity $\chi$, defined as
\begin{align}
\chi = \frac{d\Gamma^{(1)}_{\varphi \to \bar{\psi}\psi h_{ab}}/dE_l}{\Gamma^{(0)}_{\varphi \to \bar{\psi}\psi}} \, .
\end{align}
To quantify the one-loop contribution to the GW signal, we introduce the ratio $\chi_{\rm tree+one\text{-}loop}/\chi_{\rm tree}$, as defined in Eq.~(\ref{RatioRegardingChi}). This ratio depends on the renormalization scale $x = u/m_{\psi}$, as well as on the mass ratios $y = m_{\psi}/M$ and $\lambda = M/M_{\rm Pl}$.
\begin{figure}[ht]
 	\begin{center}
 	\includegraphics[scale=0.81]{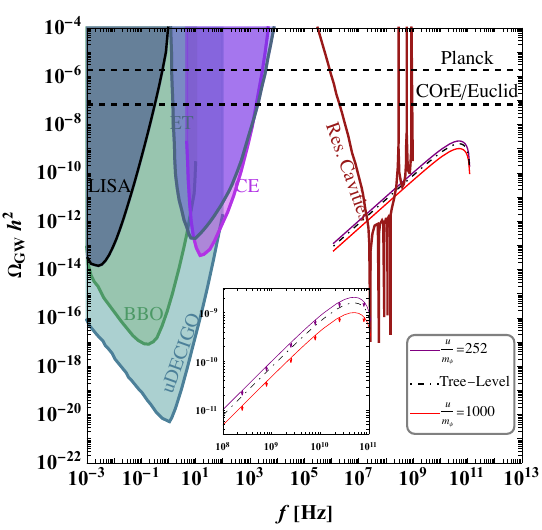}
      \includegraphics[scale=0.81]{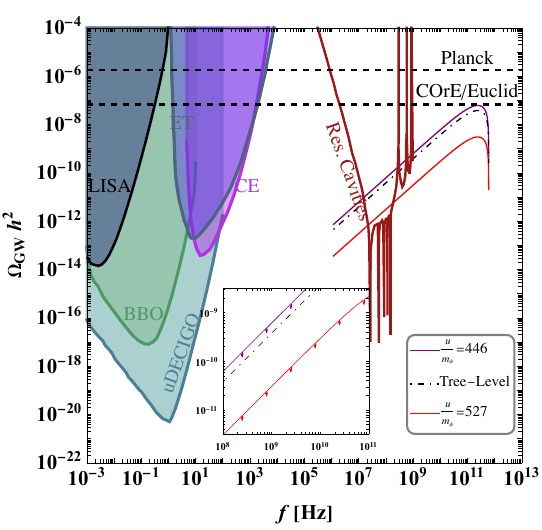}
       \caption{Gravitational-wave spectrum $\Omega_{\rm GW} h^2$ as a function of frequency $f$. The curves correspond to inflaton decay into fermion pairs at tree level (purple dot-dashed), including one-loop corrections with maximal (purple solid) and minimal (red solid) contributions. Shaded regions denote the projected sensitivity ranges of various GW detectors, while black dashed lines indicate current and future CMB constraints. 
Left panel: $\lambda=0.1$, $T_{\rm rh}=5.5\times10^{15}\,\mathrm{GeV}$, and $y=10^{-3}$. 
Right panel: $\lambda=0.4$, $T_{\rm rh}=5.5\times10^{15}\,\mathrm{GeV}$, and $y=10^{-3}$. In addition, in the left panel, the red and purple solid curves correspond to the values of $x$ at which the maximum and minimum occur, respectively, in the right panel of Fig.~\ref{MaxDeviationRenorScale} for $M/M_{\text{Pl}}=0.1$ and $y=10^{-3}$. Similarly, in the right panel, the red and purple solid curves indicate the values of $x$ corresponding to the maximum and minimum, respectively, in the right panel of Fig.~\ref{MaxDeviationRenorScaleRelaxPertur} for $M/M_{\text{Pl}}=0.4$ and $y=10^{-3}$.}
 		\label{LoopVariationInGWs}
 	\end{center}
 \end{figure}

We consider two benchmark values, $\lambda = 0.01$ and $\lambda = 0.1$, as shown in Fig.~\ref{MaxDeviationRenorScale}. We find that the ratio can reach a maximum value of $\mathcal{O}(1.4)$ for $\lambda = 0.1$, depending on the choice of $x$ and $y$, while it can also be suppressed down to $\mathcal{O}(0.6)$.
As an illustrative benchmark, we focus on the case $\lambda = 0.1$ with $y = 10^{-3}$.  We take $x = 252$ (maximal enhancement) and $x = 1000$ (minimal value), and present the resulting GW spectrum $\Omega_{\rm GW} h^2$ as a function of frequency $f$ in the left of Fig.~\ref{LoopVariationInGWs}. The projected sensitivities of GW detectors are indicated by shaded regions \cite{LISA:2017pwj,Harry:2006fi,Seto:2001qf,Reitze:2019iox,Armengaud:2014gea}, while the BBN constraints \cite{Yeh:2022heq} are shown as black dashed lines.
 
In addition, the loop correction is sensitive to the inflaton mass. For instance, taking $m_{\psi} = 0.4\,M_{\rm Pl}$, the quantity $\chi$ can be suppressed by up to two orders of magnitude, as illustrated in Fig.~\ref{MaxDeviationRenorScaleRelaxPertur}.
To further demonstrate this effect, we consider a second benchmark with $\lambda = 0.4$ and $y = 10^{-3}$. In this case, we take $x = 446$ and $x = 527$, corresponding to the maximal and minimal contributions, respectively, and present the resulting GW spectrum $\Omega_{\rm GW} h^2(f)$ in the right panel of Fig.~\ref{LoopVariationInGWs}.

\section{Conclusion and discussion \label{ConAndDiscuss}}

In this work, we investigate the one-loop corrections arising from the torsion-induced four-fermion interaction to the inflaton three-body decay processes that produce a pair of final-state fermions through the standard Yukawa interaction. After evaluating the fermion loop integrals using dimensional regularization and the minimal subtraction scheme ($\overline{\text{MS}}$), the renormalization scale remains in the logarithmic terms. Consequently, it enters both the inflaton two-body decay rate with a fermion pair in the final state and the three-body decay rate involving a fermion pair together with graviton emission. Our primary goal is to establish a connection with the stochastic gravitational-wave signal generated during inflaton decay, which is closely related to the quantity $\chi=\frac{d\Gamma_{\varphi\to\bar{\psi}\psi h_{ab}}^{(1)}/dE_{l}}{\Gamma_{\varphi\to\bar{\psi}\psi}^{(0)}}$, characterizing the ratio of the differential three-body decay width to the corresponding two-body decay width. Imposing the validity of the perturbative expansion, we therefore examine how $\chi$, including loop corrections, deviates from its tree-level value as the renormalization scale $u$ varies. 

A notable feature that emerges is the asymmetric impact of the running scale on $\chi$. As the renormalization scale $u$ varies, the loop corrections can enhance $\chi_{\text{tree+one-loop}}/\chi_{\text{tree}}$ by at most a factor of order $\mathcal{O}(1.5)$. In contrast, the suppressive effect is significantly more pronounced, with $\chi$ reduced by roughly half an order of magnitude and, in some cases, up to two orders of magnitude. Although the most substantial suppression occurs in regions where the strict applicability of perturbation theory is partially relaxed, it is nevertheless noteworthy that, under comparable conditions, the enhancement induced by the running of $u$ remains modest and does not exceed the $\mathcal{O}(1)$ level. 

These findings have potentially important implications for phenomenological models of stochastic gravitational-wave production from inflaton decay. In particular, for scenarios in which the final states consist of fermion pairs, especially those predicting observable signals or exhibiting significant spectral amplification, such as \cite{Bernal:2023wus,Kanemura:2023pnv,Lee:2026fpg}, our results suggest that loop corrections, and in particular fermionic self-interactions of the type considered here, may induce substantial suppressions of the gravitational-wave amplitude. This effect could shift the predicted signal away from the parameter regions accessible to future observations, and therefore must be taken into account in realistic model building.
\begin{figure}[H]
	\begin{center}
		\includegraphics[width=0.85\columnwidth]{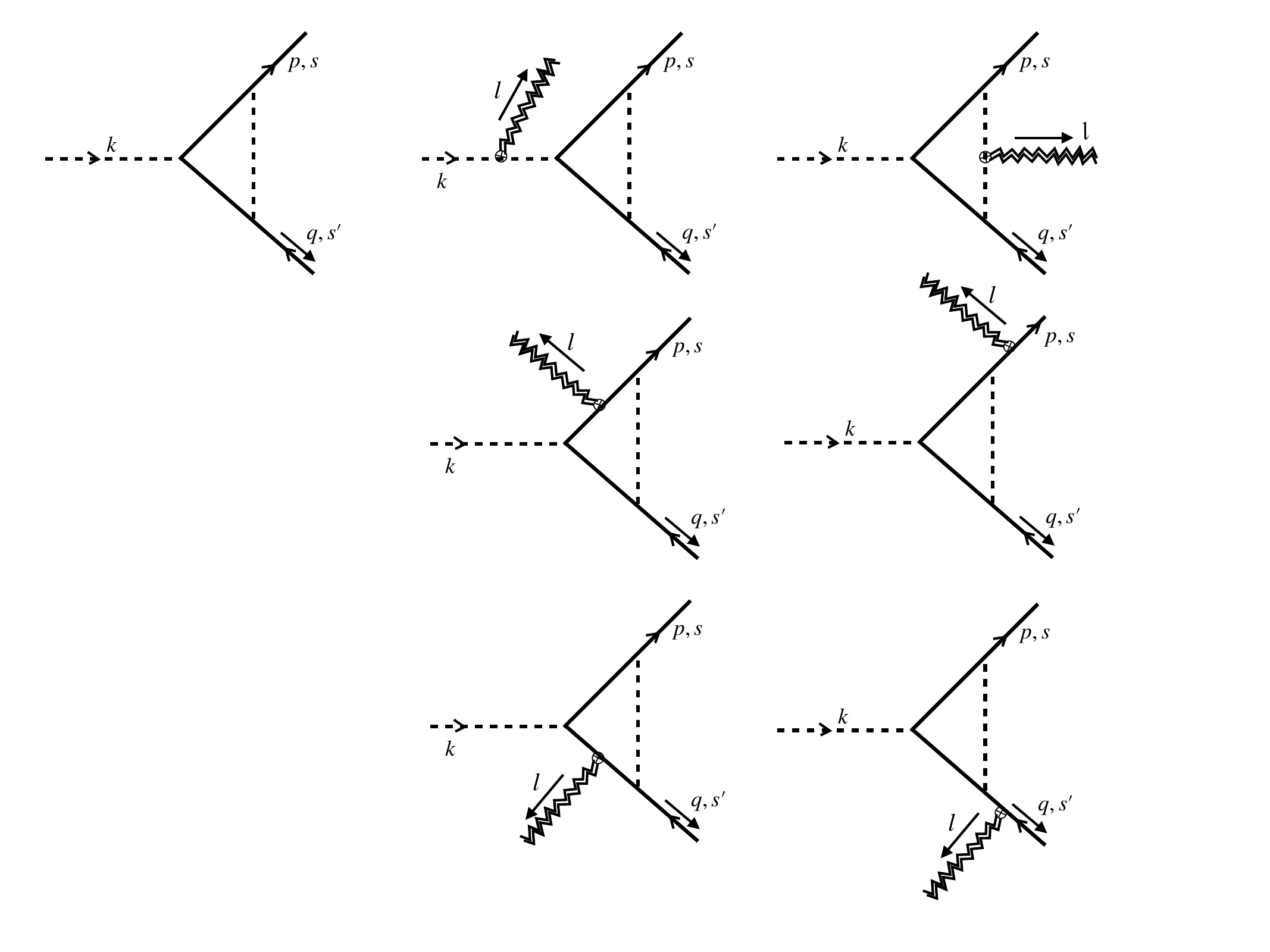}
		\caption{Feynman diagrams illustrating the one-loop corrections to the inflaton two-body and three-body decay processes within a Yukawa-interaction framework.}
		\label{YukawaOneLoopTwoThreeBody}
	\end{center}
\end{figure}

Furthermore, existing studies of inflaton three-body decays at the tree level generally indicate that the stochastic gravitational-wave signals arising from fermionic and scalar final states are nearly indistinguishable. Our results suggest that this degeneracy may be lifted once loop corrections are included, potentially leading to observable differences between the two channels. A complete assessment of this effect requires incorporating loop contributions in the scalar channel as well, for instance those arising from interactions such as $\lambda \phi^4$, which we leave for future work. In addition, the implications of loop-induced effects are not limited to decay processes. Similar considerations apply to inflaton annihilation into fermion pairs, where loop corrections, particularly those associated with torsion-induced four-fermion interactions, may again lead to significant suppression of the resulting gravitational-wave signal. A detailed analysis of these effects in annihilation channels will be left in future investigations.

As a natural extension of the above considerations, it is important to assess the regime of validity of the perturbative framework itself. In particular, when the inflaton mass $M$ approaches the Planck scale $M_{\text{pl}}$, the torsion-induced four-fermion interaction drives the system toward a strong-coupling regime, rendering the perturbative quantum field theory description unreliable. This not only motivates the development of appropriate non-perturbative approaches, but also raises a physically significant question regarding the interpretation of our results in this regime. As indicated in Fig.~\ref{MaxDeviationRenorScaleRelaxPertur}, although the one-loop approximation may lose quantitative accuracy near the Planck scale, it still suggests a qualitative trend: the running of the renormalization scale leads predominantly to a suppression, rather than an enhancement, of the gravitational-wave spectrum, in some cases by up to two orders of magnitude. It is therefore crucial to determine whether this pronounced suppression reflects a genuine physical effect or instead arises as an artifact of the breakdown of perturbation theory. Clarifying this issue will require a systematic non-perturbative treatment, which we leave for future investigation.

Finally, we emphasize that, at the one-loop level, we have consistently neglected a class of Feynman diagrams proportional to the third power of the Yukawa coupling, $y_\psi^3$, as illustrated in Fig. ~\ref{YukawaOneLoopTwoThreeBody}. These contributions arise exclusively from the Yukawa interaction sector and do not interfere with the leading structures captured in our analysis. The inclusion of such $\mathcal{O}(y_\psi^3)$ diagrams would substantially complicate the parametric analysis. In particular, both in constraining the renormalization scale $u$ via $R^{(0)}$ and $R^{(1)}$, and in quantifying the deviation encoded in $\chi/\chi_{\text{tree}}$, one would be required to systematically account for the relative hierarchy between the Yukawa coupling squared $y_\psi^2$ and the other dimensionless parameters in the problem, namely $x = u / m_\psi$, $y = m_\psi / M$, and $\lambda = M / M_{\text{Pl}}$. Moreover, once the contributions from Fig.\ref{YukawaOneLoopTwoThreeBody} are incorporated, the dependence of $\chi/\chi_{\text{tree}}$ would no longer be controlled solely by the aforementioned hierarchy parameters, but would also involve the additional kinematic variable $z = E_{l} / M$. This introduces a nontrivial interplay between coupling constants and phase-space variables, thereby significantly increasing the complexity of the analysis and obscuring the dominant physical effects. For these reasons, we defer a systematic investigation of the $\mathcal{O}(y_\psi^3)$ contributions to future work. Such an extension would nevertheless be of considerable interest. In particular, within concrete frameworks such as Higgs-inflaton models, it may provide a novel avenue to constrain Standard Model Yukawa couplings $y_\psi$ through primordial gravitational-wave observables, complementing existing bounds derived from collider experiments.

\section{Acknowledgments}
AlexKen was supported by funding from the China Scholarship Council (CSC) with grant number 202008620074. Keyun Wu was also supported by the China Scholarship Council (CSC) under grant number 202206540010 and acknowledges additional support from research grants No. PID2022-126224NB-C21 and No. 2021-SGR-249 provided by the Generalitat de Catalunya.

\appendix

\section{Useful \texorpdfstring{$\gamma$}{gamma}-matrix trace identities \label{IdentityToFermiLoopIntegral}}

In evaluating the final squared amplitudes, as well as in expanding some terms appearing in fermionic loop integrals, one inevitably encounters traces of products of multiple $\gamma$-matrices. We therefore summarize here a set of useful identities and techniques for reducing such traces. A key property of the antisymmetrized products $\gamma^{m_1 m_2 \dots m_r}$ ($4 \geq r \geq 1$) is that they are traceless. The central idea is to decompose a generic product $\gamma^{m_{1}}\gamma^{m_{2}}\dots\gamma^{m_{i}}$ (with $i$ not necessarily restricted to $4$) in terms of the Clifford algebra basis $\Gamma^{\mathcal{A}}=\{I,\gamma^{m},\gamma^{m_1m_2},\gamma^{m_1m_2m_3},\gamma^{m_1m_2m_3m_4}\}$ given in \cite{Freedman:2012zz,Lee:2025lyk}. We illustrate this strategy explicitly through the following examples. The simplest case is \begin{align}
&\gamma_{m_{1}}\gamma_{m_{2}}=\eta_{m_{1}m_{2}}I+\gamma_{m_{1}m_{2}}.
\end{align}Assisted by this decomposition, it is easy to obtain
\begin{align}
&\text{Tr}(\gamma_{m_{1}}\gamma_{m_{2}})=4\eta_{m_{1}m_{2}}.
\end{align}In next step, we go ahead with the triple product $\gamma_{m_{1}}\gamma_{m_{2}}\gamma_{m_{3}}$, namely
\begin{align}
\nonumber
\gamma_{m_{1}}\gamma_{m_{2}}\gamma_{m_{3}}&=\!\eta_{m_{1}m_{2}}\gamma_{m_{3}}\!+\frac{1}{2}\gamma_{m_{1}m_{2}}\gamma_{m_{3}}\!+\frac{1}{2}(\gamma_{m_{3}}\gamma_{m_{1}m_{2}}\!-\!2\eta_{m_{1}m_{3}}\gamma_{m_{2}}\!+\!2\gamma_{m_{1}}\eta_{m_{2}m_{3}})\\
\label{DecomposeTripleProduct}
&=\gamma_{m_{1}m_{2}m_{3}}+\eta_{m_{1}m_{2}}\gamma_{m_{3}}-\eta_{m_{1}m_{3}}\gamma_{m_{2}}+\eta_{m_{2}m_{3}}\gamma_{m_{1}}\\
\label{DecomposeTripleProductV1}
&=\!-\text{i}\epsilon_{m_{1}m_{2}m_{3}n_{1}}\gamma_{5}\gamma^{n_{1}}\!+\!\eta_{m_{1}m_{2}}\gamma_{m_{3}}\!-\!\eta_{m_{1}m_{3}}\gamma_{m_{2}}\!+\!\eta_{m_{2}m_{3}}\gamma_{m_{1}}.
\end{align}According to the above result, one could recognize $\text{Tr}\big(\gamma_{m_{1}}\gamma_{m_{2}}\gamma_{m_{3}}\big)=0$. Follow this way, the quadruple product $\gamma_{m_{1}}\gamma_{m_{2}}\gamma_{m_{3}}\gamma_{m_{4}}$ is treated as
\begin{align}
\nonumber
\gamma_{m_{1}}\gamma_{m_{2}}\gamma_{m_{3}}\gamma_{m_{4}}=&\gamma_{m_{1}m_{2}m_{3}m_{4}}\!+\!\eta_{m_{1}m_{4}}\gamma_{m_{2}m_{3}}\!-\!\eta_{m_{2}m_{4}}\gamma_{m_{1}m_{3}}\!+\!\eta_{m_{3}m_{4}}\gamma_{m_{1}m_{2}}\\
\nonumber
&+\!\eta_{m_{1}m_{2}}(\gamma_{m_{3}m_{4}}\!+\!\eta_{m_{3}m_{4}}I)-\eta_{m_{1}m_{3}}(\gamma_{m_{2}m_{4}}+\eta_{m_{2}m_{4}}I)\\
\label{DecomposeFourfoldProduct}
&+\eta_{m_{2}m_{3}}(\gamma_{m_{1}m_{4}}+\eta_{m_{1}m_{4}}I).
\end{align}Basing on the expression Eq. \eqref{DecomposeFourfoldProduct}, it is distinct to give
\begin{align}
\label{TraceQuadProduct}
&\text{Tr}(\gamma_{m_{1}}\gamma_{m_{2}}\gamma_{m_{3}}\gamma_{m_{4}})=4(\eta_{m_{1}m_{2}}\eta_{m_{3}m_{4}}-\eta_{m_{1}m_{3}}\eta_{m_{2}m_{4}}+\eta_{m_{1}m_{4}}\eta_{m_{2}m_{3}}).
\end{align}Alternatively, with the equipment of dual relations Eq. \eqref{DecomposeTripleProductV1}, it enables us to rewrite the quadruple product as
\begin{align}
\nonumber
\gamma_{m_{1}}\gamma_{m_{2}}\gamma_{m_{3}}\gamma_{m_{4}}=&-\text{i}\epsilon_{m_{1}m_{2}m_{3}n_{1}}\gamma_{5}(\gamma_{~~m_{4}}^{n_{1}}+\delta_{m_{4}}^{n_{1}}I)+\eta_{m_{1}m_{2}}(\gamma_{m_{3}m_{4}}+\eta_{m_{3}m_{4}}I)\\
&-\eta_{m_{1}m_{3}}(\gamma_{m_{2}m_{4}}+\eta_{m_{2}m_{4}}I)+\eta_{m_{2}m_{3}}(\gamma_{m_{1}m_{4}}+\eta_{m_{1}m_{4}}I).
\end{align}which also leads to the result Eq. \eqref{TraceQuadProduct}. Analogously, after repeating these procedures for many times, it is easy to observe
\begin{align}
\label{VanishOddSingleGamma}
&\text{Tr}(\gamma_{m_{1}}\gamma_{m_{2}}\dots\dots\gamma_{m_{2i+1}})=0.
\end{align}After reviewing some basic properties of traces involving $\gamma$-matrices and working through several illustrative examples, we now present a set of identities relevant for evaluating more involved trace structures. These identities will be used in the computation of the one-loop Feynman diagrams shown in Fig.~\ref{TorsionOneLoopTwoBody} and Fig.~\ref{TorsionOneLoopThreeBodyGWs}. We begin with the identities required for evaluating the one-loop corrected two-body inflaton decay, \begin{align}
\nonumber
&\text{Tr}\big((\cancel{\tilde{k}}+\cancel{\tilde{p}}+m_{\psi}I)\gamma^{mna}(\cancel{\tilde{k}}+m_{\psi}I)\big)\xrightarrow{\gamma^{m_{1}m_{2}m_{3}}=-\text{i}\epsilon^{m_{1}m_{2}m_{3}m_{4}}\gamma_{5}\gamma_{m_{4}}\,,\,\text{Tr}(\gamma^{mna})=0}\\
\nonumber
&=\text{i}\epsilon^{mnaa_{1}}\text{Tr}\big((\tilde{p}\!+\!\tilde{k})^{b_{1}}\!\tilde{k}^{c_{1}}\!\gamma_{5}(\gamma_{b_{1}a_{1}}\!+\!\eta_{b_{1}a_{1}}I)\gamma_{c_{1}}\!+\!m_{\psi}(\tilde{p}\!+\!\tilde{k})^{c_{1}}\gamma_{5}(\gamma_{c_{1}a_{1}}\!+\!\eta_{c_{1}a_{1}}I)\!-\!m_{\psi}\tilde{k}^{c_{1}}\!\gamma_{5}(\gamma_{a_{1}c_{1}}\!+\!\eta_{a_{1}c_{1}}I)\big)\\
\nonumber
&\quad\xrightarrow{\gamma_{5}\gamma_{m_{1}m_{2}}\propto\epsilon_{m_{1}m_{2}m_{3}m_{4}}\gamma^{m_{4}m_{3}}\,,\,\gamma_{5}\gamma_{m_{1}}\propto\epsilon_{m_{1}m_{2}m_{3}m_{4}}\gamma^{m_{4}m_{3}m_{2}}\Rightarrow\text{Tr}(\gamma_{5}\gamma_{m_{1}m_{2}})=\text{Tr}(\gamma_{5}\gamma_{m_{1}})=0\,,\,\text{Tr}(\gamma_{5})=0}\\
\nonumber
&=\text{i}\epsilon^{mnaa_{1}}(\tilde{p}+\tilde{k})^{b_{1}}\tilde{k}^{c_{1}}\text{Tr}(\gamma_{5}\gamma_{b_{1}a_{1}}\gamma_{c_{1}})=\frac{\text{i}}{2}\epsilon^{mnaa_{1}}(\tilde{p}+\tilde{k})^{b_{1}}\tilde{k}^{c_{1}}\text{Tr}(\gamma_{5}\gamma_{b_{1}a_{1}}\cdot\gamma_{c_{1}}+\gamma_{c_{1}}\cdot\gamma_{5}\gamma_{b_{1}a_{1}})\\
\nonumber
&=\frac{\text{i}}{2}\epsilon^{mnaa_{1}}(\tilde{p}+\tilde{k})^{b_{1}}\tilde{k}^{c_{1}}\text{Tr}\big(\gamma_{5}(\gamma_{b_{1}a_{1}}\gamma_{c_{1}}-\gamma_{c_{1}}\gamma_{b_{1}a_{1}})\big)\\
\label{TraceIdentiTwoBodyLoop}
&=\text{i}\epsilon^{mnaa_{1}}(\tilde{p}+\tilde{k})^{b_{1}}\tilde{k}^{c_{1}}\underbrace{\text{Tr}(\eta_{a_{1}c_{1}}\gamma_{5}\gamma_{b_{1}}-\eta_{b_{1}c_{1}}\gamma_{5}\gamma_{a_{1}})}_{\text{Tr}(\gamma_{5}\gamma_{m_{1}})=0}=0.
\end{align}This identity will also be employed in the computation of the one-loop corrected three-body decay processes corresponding to diagrams 1, 2, and 4 in Fig.~\ref{TorsionOneLoopThreeBodyGWs}. Furthermore, from the derivation of the identity in Eq. \eqref{TraceIdentiTwoBodyLoop}, one readily obtains \begin{align}
\label{VanishTraceMultipleGamma}
&\text{Tr}(\gamma^{c_{1}}\gamma^{mna})=\text{Tr}(\gamma^{c_{2}}\gamma^{c_{1}}\gamma^{mna})=\text{Tr}(\gamma^{c_{2}c_{1}}\gamma^{mna})=0.
\end{align}Based on this result, one can further derive
\begin{align}
\nonumber
&\text{Tr}(\gamma^{mna}\gamma^{c_{1}}\gamma^{c_{2}}\gamma^{c_{3}})\!=\!\text{Tr}(\gamma^{mna}\gamma^{c_{1}c_{2}}\gamma^{c_{3}})\!=\!\frac{1}{2}\text{Tr}\big(\gamma^{mna}(\gamma^{c_{1}c_{2}}\gamma^{c_{3}}\!+\!2\gamma^{c_{1}}\eta^{c_{2}c_{3}}\!-\!2\eta^{c_{1}c_{3}}\gamma^{c_{2}}\!+\!\gamma^{c_{3}}\gamma^{c_{1}c_{2}})\big)\\
\label{TraceGamma3WithThreeSigGamma}
&=\text{Tr}(\gamma^{mna}\gamma^{c_{1}c_{2}c_{3}})=-\epsilon^{mnaa_{1}}\epsilon^{c_{1}c_{2}c_{3}n_{1}}\text{Tr}(\gamma_{5}\gamma_{a_{1}}\gamma_{5}\gamma_{n_{1}})=4\eta_{a_{1}n_{1}}\epsilon^{mnaa_{1}}\epsilon^{c_{1}c_{2}c_{3}n_{1}},
\end{align}and
\begin{align}
\nonumber
&\text{Tr}(\gamma^{mna}\gamma^{c_{1}}\gamma^{c_{2}}\gamma^{c_{3}}\gamma^{c_{4}})\!=\!\text{Tr}(\gamma^{mna}\gamma^{c_{1}c_{2}}\gamma^{c_{3}}\gamma^{c_{4}})\!=\!\text{Tr}\big(\gamma^{mna}(\gamma^{c_{3}c_{1}c_{2}}\gamma^{c_{4}}\!+\!\gamma^{c_{3}}\gamma^{c_{4}}\eta^{c_{2}n_{1}}\!-\!\eta^{c_{1}c_{3}}\gamma^{c_{2}}\gamma^{c_{4}})\big)\\
\nonumber
&=\text{Tr}\big(\gamma^{mna}\gamma^{c_{3}c_{1}c_{2}}\gamma^{c_{4}}\big)=\text{Tr}\big(\gamma^{mna}(\gamma^{c_{4}c_{1}}\eta^{c_{2}c_{3}}\!-\!\gamma^{c_{4}c_{2}}\eta^{c_{1}c_{3}}\!+\!\gamma^{c_{1}c_{2}}\eta^{c_{4}c_{3}}\!-\!\gamma^{c_{3}c_{4}c_{1}c_{2}})\big)\\
\label{TraceGamma3WithFourSigGamma}
&=-\text{Tr}\big(\gamma^{mna}\gamma^{c_{1}c_{2}c_{3}c_{4}}\big)\propto\epsilon^{mnaa_{1}}\epsilon^{c_{1}c_{2}c_{3}c_{4}}\text{Tr}(\gamma_{5}\gamma_{a_{1}}\gamma_{5})=0.
\end{align}In fact, the result in Eq. \eqref{TraceGamma3WithFourSigGamma} can alternatively be obtained from the identity Eq. \eqref{VanishOddSingleGamma}. By employing the trace identities Eq. \eqref{VanishTraceMultipleGamma}-- \eqref{TraceGamma3WithFourSigGamma}, we can simplify the following expressions relevant for the Feynman diagrams labeled 3 and 5 in Fig.~\ref{TorsionOneLoopThreeBodyGWs},
\begin{align}
\nonumber
&\text{Tr}\big(\gamma^{mna}(\cancel{q}^{\prime}+m_{\psi}I)(\cancel{q}^{\prime}-\cancel{k}+m_{\psi}I)\gamma^{n_{1}}(\cancel{q}^{\prime}-\cancel{k}+\cancel{l}+m_{\psi}I)\big)\times\bar{u}(\boldsymbol{p},s)\gamma_{mna}v(\boldsymbol{q},s^{\prime})\\
\nonumber
&=4m_{\psi}(k-2q^{\prime})_{c_{1}}l_{c_{2}}\eta_{c_{3}a_{1}}\epsilon^{n_{1}c_{1}c_{2}c_{3}}\epsilon^{mnaa_{1}}\times\bar{u}(\boldsymbol{p},s)\gamma_{mna}v(\boldsymbol{q},s^{\prime})\\
\label{TraceThreeBodyOneLoopNum3}
&=-24m_{\psi}(k-2q^{\prime})_{c_{1}}l_{c_{2}}\times\bar{u}(\boldsymbol{p},s)\gamma^{c_{1}c_{2}n_{1}}v(\boldsymbol{q},s^{\prime}),
\end{align}and
\begin{align}
\nonumber
&\text{Tr}\big(\gamma^{mna}(\cancel{q}^{\prime}-\cancel{l}+m_{\psi}I)\gamma^{n_{1}}(\cancel{q}^{\prime}+m_{\psi}I)(\cancel{q}^{\prime}-\cancel{k}+m_{\psi}I)\big)\times\bar{u}(\boldsymbol{p},s)\gamma_{mna}v(\boldsymbol{q},s^{\prime})\\
\nonumber
&=4m_{\psi}(k_{c_{1}}-2q_{c_{1}}^{\prime})l_{c_{2}}\eta_{c_{3}a_{1}}\epsilon^{n_{1}c_{1}c_{2}c_{3}}\times\bar{u}(\boldsymbol{p},s)\epsilon^{mnaa_{1}}\gamma_{mna}v(\boldsymbol{q},s^{\prime})\\
\label{TraceThreeBodyOneLoopNum5}
&=-24m_{\psi}(k-2q^{\prime})_{c_{1}}l_{c_{2}}\times\bar{u}(\boldsymbol{p},s)\gamma^{c_{1}c_{2}n_{1}}v(\boldsymbol{q},s^{\prime}).
\end{align}In deriving these results, we repeatedly make use of the expansion
\begin{align}
\nonumber
(\cancel{q}^{\prime}+\cancel{A}+m_{\psi}I)(\cancel{q}^{\prime}+\cancel{B}+m_{\psi}I)&=(q^{\prime2}+2q^{\prime}\cdot B)I+(\cancel{A}-\cancel{B}+2m_{\psi}I)\cancel{q}^{\prime}\\
&+\cancel{A}\cancel{B}+m_{\psi}(\cancel{B}+\cancel{A}+m_{\psi}I).
\end{align}

\section{Explicit expressions for the one-loop corrected two-body and three-body decay amplitudes \label{ThreeBodyAmpliOneLoop}}

The complete scattering amplitudes corresponding to Fig.~\ref{TorsionOneLoopTwoBody} can be written explicitly as
\begin{small}
\begin{align}
\nonumber
\mathcal{M}_{\varphi(k)\to\bar{\psi}(p,s)\psi(q,s^{\prime})}^{\text{(tree+one-loop)}}\!=&\!-\text{i}y_{\psi}\bar{u}(\boldsymbol{p},s)v(\boldsymbol{q},s^{\prime})\!+\!\frac{y_{\psi}\kappa^{2}}{32}\!\int \!\! d^{4}q^{\prime}\bar{u}(\boldsymbol{p},s)\\
\nonumber
&\times\big\{\text{Tr}\big(\frac{\text{i}}{(\cancel{q}^{\prime}-\cancel{k})-m_{\psi}I+\text{i}\epsilon}\gamma^{mna}\frac{\text{i}}{\cancel{q}^{\prime}-m_{\psi}I+\text{i}\epsilon}\big)\gamma_{mna}\\
\label{AmplitudeTwoBodyOneLoopv1}
&-3!\big(\gamma_{5}\gamma_{a}\frac{\text{i}}{\cancel{q^{\prime}}-m_{\psi}I+\text{i}\epsilon}\frac{\text{i}}{(\cancel{q}^{\prime}-\cancel{k})-m_{\psi}I+\text{i}\epsilon}\gamma_{5}\gamma^{a}\big)\big\} v(\boldsymbol{q},s^{\prime}),
\end{align}
\end{small}where the first term in the fermion loop integral vanishes, as shown in Eq. \eqref{TraceIdentiTwoBodyLoop}. After performing the loop integrals using dimensional regularization and the minimal subtraction scheme ($\overline{\text{MS}}$), we arrive at the result given in Eq. \eqref{AmplitudeTwoBodyOneLoopv2}.

On the other hand, the complete scattering amplitudes for the three-body decay process, as depicted in Fig.~\ref{TorsionOneLoopThreeBodyGWs}, can be expressed as
\begin{align}
\nonumber
\mathcal{M}_{\varphi(k)\to\bar{\psi}(p,s)\psi(q,s^{\prime})h_{m_{1}n_{1}}(l,\chi)}^{\text{(tree)}}&=\text{i}\kappa y_{\psi}e_{m_{1}n_{1}}(\boldsymbol{l},\chi)\big\{\frac{k^{m_{1}}k^{n_{1}}}{(k-l)^{2}\!-\!M^{2}}\bar{u}(\boldsymbol{p},s)v(\boldsymbol{q},s^{\prime})\\
\nonumber
&-\frac{1}{4}\bar{u}(\boldsymbol{p},s)(p^{m_{1}}\gamma^{n_{1}}+p^{n_{1}}\gamma^{m_{1}})\frac{1}{\cancel{p}+\cancel{l}-m_{\psi}I}v(\boldsymbol{q},s^{\prime})\\
\label{ThreeBodyDecayTree}
&-\frac{1}{4}\bar{u}(\boldsymbol{p},s)\frac{1}{\cancel{q}+\cancel{l}+m_{\psi}I}(q^{m_{1}}\gamma^{n_{1}}+q^{n_{1}}\gamma^{m_{1}})v(\boldsymbol{q},s^{\prime})\big\},
\end{align}and
\begin{align}
\label{BriefOneLoopExpress}
&\mathcal{M}_{\varphi(k)\to\bar{\psi}(p,s)\psi(q,s^{\prime})h_{m_{1}n_{1}}(l,\chi)}^{\text{(one-loop)}}=\sum_{n=1}^{5}\mathcal{M}_{\varphi(k)\to\bar{\psi}(p,s)\psi(q,s^{\prime})h_{m_{1}n_{1}}(l,\chi)}^{\text{(one-loop)},(n)},
\end{align}
where the superscript $(n)$ labels the individual one-loop diagrams and corresponds directly to the numbering in Fig.~\ref{TorsionOneLoopThreeBodyGWs}. Each term in Eq. \eqref{BriefOneLoopExpress} can be explicitly expanded as
\begin{align}
\nonumber
\mathcal{M}_{\varphi(k)\to\bar{\psi}(p,s)\psi(q,s^{\prime})h_{m_{1}n_{1}}(l,\chi)}^{\text{(one-loop)},(1)}&=\!-\frac{y_{\psi}\kappa^{3}k^{m_{1}}k^{n_{1}}e_{m_{1}n_{1}}(\boldsymbol{l},\chi)}{64k\cdot l}\int\frac{d^{4}q^{\prime}}{(2\pi)^{4}}\frac{1}{(q^{\prime2}-m_{\psi}^{2})\big((q^{\prime}-k+l)^{2}-m_{\psi}^{2}\big)}\\
\nonumber
&\times\big\{\underbrace{\text{Tr}\big((\cancel{q}^{\prime}\!-\!\cancel{k}\!+\!\cancel{l}\!+\!m_{\psi}I)\gamma^{mna}(\cancel{q}^{\prime}\!+\!m_{\psi}I)\big)\bar{u}(\boldsymbol{p},s)\gamma_{mna}v(\boldsymbol{q},s^{\prime})}_{\text{vanish}}\\ \label{ThreeBodyLoopAmplitude1}
&+6\bar{u}(\boldsymbol{p},s)\gamma_{c_{1}}(\cancel{q}^{\prime}\!-\!m_{\psi})(\cancel{q}^{\prime}\!-\!\cancel{k}\!+\!\cancel{l}\!-\!m_{\psi})\gamma^{c_{1}}v(\boldsymbol{q},s^{\prime})\big\},
\end{align}
\begin{align}
\nonumber
\mathcal{M}_{\varphi(k)\to\bar{\psi}(p,s)\psi(q,s^{\prime})h_{m_{1}n_{1}}(l,\chi)}^{\text{(one-loop)},(2)}&=\!-\frac{y_{\psi}\kappa^{3}e_{m_{1}n_{1}}(\boldsymbol{l},\chi)}{64}\int\frac{d^{4}q^{\prime}}{(2\pi)^{4}}\frac{(q^{\prime}-k)^{m_{1}}}{(q^{\prime2}-m_{\psi}^{2})\big((q^{\prime}-k)^{2}-m_{\psi}^{2}\big)\big((q^{\prime}-k+l)^{2}-m_{\psi}^{2}\big)}\\ \label{ThreeBodyLoopAmplitude2}
\nonumber
&\times\big\{\underbrace{\text{Tr}\big(\gamma^{mna}(\cancel{q}^{\prime}\!+\!m_{\psi}I)(\cancel{q}^{\prime}\!-\!\cancel{k}\!+\!m_{\psi}I)\gamma^{n_{1}}(\cancel{q}^{\prime}\!-\!\cancel{k}\!+\!\cancel{l}\!+\!m_{\psi}I)\big)\bar{u}(\boldsymbol{p},s)\gamma_{mna}v(\boldsymbol{q},s^{\prime})}_{-24m_{\psi}(k-2q^{\prime})_{c_{1}}l_{c_{2}}\times\bar{u}(\boldsymbol{p},s)\gamma^{c_{1}c_{2}n_{1}}v(\boldsymbol{q},s^{\prime})}\\
&+6\bar{u}(\boldsymbol{p},s)\gamma_{c_{1}}(\cancel{q}^{\prime}\!-\!m_{\psi}I)\big((\cancel{q}^{\prime}\!-\!\cancel{k})\!-\!m_{\psi}I\big)\gamma^{n_{1}}(\cancel{q}^{\prime}\!-\!\cancel{k}\!+\!\cancel{l}\!-\!m_{\psi}I)\gamma^{c_{1}}v(\boldsymbol{q},s^{\prime})\big\},
\end{align}
\begin{align}
\nonumber
\mathcal{M}_{\varphi(k)\to\bar{\psi}(p,s)\psi(q,s^{\prime})h_{m_{1}n_{1}}(l,\chi)}^{\text{(one-loop)},(3)}&=\!-\frac{y_{\psi}\kappa^{3}p^{m_{1}}e_{m_{1}n_{1}}(\boldsymbol{l},\chi)}{64\big((p+l)^{2}-m_{\psi}^{2}\big)}\int\frac{d^{4}q^{\prime}}{(2\pi)^{4}}\frac{1}{(q^{\prime2}-m_{\psi}^{2})\big((q^{\prime}-k)^{2}-m_{\psi}^{2}\big)}\\ \label{ThreeBodyLoopAmplitude3}
\nonumber
&\times\big\{\underbrace{\text{Tr}\big((\cancel{q}^{\prime}\!-\!\cancel{k}\!+\!m_{\psi}I)\gamma^{mna}(\cancel{q}^{\prime}\!+\!m_{\psi}I)\big)\bar{u}(\boldsymbol{p},s)\gamma^{n_{1}}(\cancel{p}\!+\!\cancel{l}\!+\!m_{\psi}I)\gamma_{mna}v(\boldsymbol{q},s^{\prime})}_{\text{vanish}}\\
&+6\bar{u}(\boldsymbol{p},s)\gamma^{n_{1}}(\cancel{p}\!+\!\cancel{l}\!+\!m_{\psi}I)\gamma_{c_{1}}(\cancel{q}^{\prime}\!-\!m_{\psi}I)((\cancel{q}^{\prime}\!-\!\cancel{k})\!-\!m_{\psi}I)\gamma^{c_{1}}v(\boldsymbol{q},s^{\prime})\big\},
\end{align}
\begin{align}
\nonumber
\mathcal{M}_{\varphi(k)\to\bar{\psi}(p,s)\psi(q,s^{\prime})h_{m_{1}n_{1}}(l,\chi)}^{\text{(one-loop)},(4)}&=\!-\frac{y_{\psi}\kappa^{3}e_{m_{1}n_{1}}(\boldsymbol{l},\chi)}{64}\int\frac{d^{4}q^{\prime}}{(2\pi)^{4}}\frac{q^{\prime m_{1}}}{(q^{\prime2}-m_{\psi}^{2})\big((q^{\prime}-k)^{2}-m_{\psi}^{2}\big)\big((q^{\prime}-l)^{2}-m_{\psi}^{2}\big)}\\ \label{ThreeBodyLoopAmplitude4}
\nonumber
&\times\big\{\underbrace{\text{Tr}\big(\gamma^{mna}(\cancel{q}^{\prime}\!-\!\cancel{l}\!+\!m_{\psi}I)\gamma^{n_{1}}(\cancel{q}^{\prime}\!+\!m_{\psi}I)(\cancel{q}^{\prime}\!-\!\cancel{k}\!+\!m_{\psi}I)\big)\bar{u}(\boldsymbol{p},s)\gamma_{mna}v(\boldsymbol{q},s^{\prime})}_{-24m_{\psi}(k_{c_{1}}-2q_{c_{1}}^{\prime})l_{c_{2}}\times\bar{u}(\boldsymbol{p},s)\gamma^{n_{1}c_{1}c_{2}}v(\boldsymbol{q},s^{\prime})}\\
&+6\bar{u}(\boldsymbol{p},s)\gamma_{c_{1}}\big((\cancel{q}^{\prime}\!-\!\cancel{l})\!-\!m_{\psi}I\big)\gamma^{n_{1}}(\cancel{q}^{\prime}\!-\!m_{\psi}I)\big((\cancel{q}^{\prime}\!-\!\cancel{k})\!-\!m_{\psi}I\big)\gamma^{c_{1}}v(\boldsymbol{q},s^{\prime})\big\},
\end{align}
\begin{align}
\nonumber
\mathcal{M}_{\varphi(k)\to\bar{\psi}(p,s)\psi(q,s^{\prime})h_{m_{1}n_{1}}(l,\chi)}^{\text{(one-loop)},(5)}&=\!-\frac{y_{\psi}\kappa^{3}q^{m_{1}}e_{m_{1}n_{1}}(\boldsymbol{l},\chi)}{64((q\!+\!l)^{2}\!-\!m_{\psi}^{2})}\int\frac{d^{4}q^{\prime}}{(2\pi)^{4}}\frac{1}{(q^{\prime2}\!-\!m_{\psi}^{2})((q^{\prime}\!-\!k)^{2}\!-\!m_{\psi}^{2})}\\ \label{ThreeBodyLoopAmplitude5}
\nonumber
&\times\big\{\underbrace{\text{Tr}\big((\cancel{q}^{\prime}\!-\!\cancel{k}\!+\!m_{\psi}I)\gamma^{mna}(\cancel{q}^{\prime}\!+\!m_{\psi}I)\big)\bar{u}(\boldsymbol{p},s)\gamma_{mna}(\cancel{q}\!+\!\cancel{l}\!-\!m_{\psi}I)\gamma^{n_{1}}v(\boldsymbol{q},s^{\prime})}_{\text{vanish}}\\
&+6\bar{u}(\boldsymbol{p},s)\gamma_{c_{1}}(\cancel{q}^{\prime}-m_{\psi}I)(\cancel{q}^{\prime}-\cancel{k}-m_{\psi}I)\gamma^{c_{1}}(\cancel{q}+\cancel{l}-m_{\psi}I)\gamma^{n_{1}}v(\boldsymbol{q},s^{\prime})\big\}.
\end{align}where the identities \eqref{TraceIdentiTwoBodyLoop} and \eqref{TraceThreeBodyOneLoopNum3}--\eqref{TraceThreeBodyOneLoopNum3} have been employed to evaluate the terms proportional to the traces in each amplitude. After evaluating the loop integrals using dimensional regularization and the minimal subtraction scheme ($\overline{\text{MS}}$), we obtain
\begin{align}
\nonumber
\mathcal{M}_{\varphi(k)\to\bar{\psi}(p,s)\psi(q,s^{\prime})h_{m_{1}n_{1}}(l,\chi)}^{\text{(one-loop)},(1)}\big\vert_{\overline{\text{MS}}}&=\!-\frac{3y_{\psi}\kappa^{3}k^{m_{1}}k^{n_{1}}e_{m_{1}n_{1}}(\boldsymbol{l},\chi)}{32k\cdot l}\bar{u}(\boldsymbol{p},s)v(\boldsymbol{q},s^{\prime})\times\big\{(4k\cdot l+8m_{\psi}^{2}-2M^{2})F_{2}\\ 
&\hspace{-1.1cm}+2\log(\frac{u^{2}}{m_{\psi}^{2}})(2k\cdot l+6m_{\psi}^{2}-M^{2})+6k\cdot l+14m_{\psi}^{2}-3M^{2}\big\},
\end{align}
\begin{align}
\nonumber
\mathcal{M}_{\varphi(k)\to\bar{\psi}(p,s)\psi(q,s^{\prime})h_{m_{1}n_{1}}(l,\chi)}^{\text{(one-loop)},(3)}\big\vert_{\overline{\text{MS}}}&=\!\frac{3\text{i}y_{\psi}\kappa^{3}p^{m_{1}}e_{m_{1}n_{1}}(\boldsymbol{l},\chi)}{32\big(m_{\psi}^{2}-(p+l)^{2}\big)}\bar{u}(\boldsymbol{p},s)(\gamma^{n_{1}}\cancel{l}+2p^{n_{1}})v(\boldsymbol{q},s')\\ 
&\hspace{-1cm}\times\big\{2m_{\psi}^{2}(4F_{1}+6\log(\frac{u^{2}}{m_{\psi}^{2}})+7)-M^{2}(2F_{1}+2\log(\frac{u^{2}}{m_{\psi}^{2}})+3)\big\},
\end{align}
\begin{align}
\nonumber
\mathcal{M}_{\varphi(k)\to\bar{\psi}(p,s)\psi(q,s^{\prime})h_{m_{1}n_{1}}(l,\chi)}^{\text{(one-loop)},(5)}\big\vert_{\overline{\text{MS}}}&=\frac{-3\text{i}y_{\psi}\kappa^{3}q^{m_{1}}e_{m_{1}n_{1}}(\boldsymbol{l},\chi)}{32\big(m_{\psi}^{2}-(q+l)^{2}\big)}\bar{u}(\boldsymbol{p},s)(\gamma^{n_{1}}\cancel{l}-2q^{n_{1}})v(\boldsymbol{q},s')\\ 
&\hspace{-1cm}\times\big\{2m_{\psi}^{2}\big(4F_{1}+6\log(\frac{u^{2}}{m_{\psi}^{2}})+7\big)-M^{2}\big(2F_{1}+2\log(\frac{u^{2}}{m_{\psi}^{2}})+3\big)\big\},
\end{align}
\begin{align}
\mathcal{M}_{\varphi(k)\to\bar{\psi}(p,s)\psi(q,s^{\prime})h_{m_{1}n_{1}}(l,\chi)}^{\text{(one-loop)},(2)}\big\vert_{\overline{\text{MS}}}&=\frac{y_{\psi}\kappa^3k^{n_1}e_{m_1,n_1}(\boldsymbol{l},x)}{192E_l^2M^3(2E_l-M)}\\ \nonumber
&\hspace{-1cm}\times\bigg[k^{m_1}\bigg(M_1\bar{u}(\boldsymbol{p},s)\cancel{l}v(\boldsymbol{q},s')+M_2\bar{u}(\boldsymbol{p},s)\cancel{k}v(\boldsymbol{q},s')+M_3\bar{u}(\boldsymbol{p},s)v(\boldsymbol{q},s')\bigg)\\ \nonumber
&\hspace{-1cm}+M_4\bar{u}(\boldsymbol{p},s)\gamma^{m_1}\cancel{l}v(\boldsymbol{q},s')
+M_5\bar{u}(\boldsymbol{p},s)\gamma^{m_1}\cancel{k}v(\boldsymbol{q},s')+M_6\bar{u}(\boldsymbol{p},s)\gamma^{m_1}v(\boldsymbol{q},s')\bigg]\\ \nonumber
&\hspace{-1cm}+\frac{y_{\psi}\kappa^3k^{m_1}e_{m_1,n_1}(\boldsymbol{l},x)}{192E_l^2M^3(2E_l-M)}\times\bigg(M_7 \bar{u}(\boldsymbol{p},s)\gamma^{n_1}\cancel{l}v(\boldsymbol{q},s')+M_8\bar{u}(\boldsymbol{p},s)\gamma^{n_1}v(\boldsymbol{q},s')\bigg),
\end{align}
\begin{align}
\mathcal{M}_{\varphi(k)\to\bar{\psi}(p,s)\psi(q,s^{\prime})h_{m_{1}n_{1}}(l,\chi)}^{\text{(one-loop)},(4)}\big\vert_{\overline{\text{MS}}}&=\frac{y_{\psi}\kappa^3k^{n_1}e_{m_1,n_1}(\boldsymbol{l},\chi)}{192E_l^2M^3(2E_l-M)}\\ \nonumber
&\hspace{-1cm}\times\bigg[k^{m_1}\bigg(M_{9}\bar{u}(\boldsymbol{p},s)\cancel{l}v(\boldsymbol{q},s')+M_{10}\bar{u}(\boldsymbol{p},s)\cancel{k}v(\boldsymbol{q},s')+M_{11}\bar{u}(\boldsymbol{p},s)v(\boldsymbol{q},s')\bigg)\\ \nonumber
&\hspace{-1cm}+M_{12}\bar{u}(\boldsymbol{p},s)\gamma^{m_1}\cancel{l}v(\boldsymbol{q},s')+M_{13}\bar{u}(\boldsymbol{p},s)\gamma^{m_1}\cancel{k}v(\boldsymbol{q},s')+M_{14}\bar{u}(\boldsymbol{p},s)\gamma^{m_1}v(\boldsymbol{q},s')\bigg]\\ \nonumber
&\hspace{-1cm}+\frac{y_{\psi}\kappa^3k^{m_1}e_{m_1,n_1}(\boldsymbol{l},x)}{192E_l^2M^3(2E_l-M)}\times\bigg(M_{15}\bar{u}(\boldsymbol{p},s)\gamma^{n_1}\cancel{l}v(\boldsymbol{q},s')+M_{16}\bar{u}(\boldsymbol{p},s)\gamma^{n_1}v(\boldsymbol{q},s')\bigg),
\end{align}
where, for notational convenience, we introduce the following shorthand quantities:
\begin{align}
&M_{1}=-6m_{\psi}M\bigg((4E_l^2M+4E_lm_{\psi}^2(-2F_1+3F_2+2)\\ \nonumber
&\hspace{1cm}+2E_lM^2(\Delta F-1)-M\Delta F(M^2-4m_{\psi}^2)\bigg)\\
&M_2=-6m_{\psi}E_l\left(2E_lM^2\Delta F+8E_lm_{\psi}^2(F_1+2)+M(M^2-4m_{\psi}^2)\Delta F\right)\\
&M_3=-6E_lM^2(2E_l-M)\bigg(E_lM(3F_2+6\log\left(\frac{u^2}{m_{\psi}^2}\right)+10)+(3M^2-12m_{\psi}^2)\Delta F\bigg)\\
&M_4=6E_l^2M^3(2E_l-M)\\
&M_5=-9E_lM^2(2E_l-M)\left(2E_lM(F_2+\log\left(\frac{u^2}{m_{\psi}^2}\right)+4)+M^2\Delta F+m_{\psi}^2\Delta L\right)\\
&M_6=2E_lm_{\psi} M^2(2E_l-M)\left(2E_lM(3F_2+3\log\left(\frac{u^2}{m_{\psi}^2}\right)+5)-12m_{\psi}^2\Delta F+3M^2\Delta F\right)\\
&M_7=3E_lM^2(2E_l-M)\left(M\left(E_l(6F_2+6\log\left(\frac{u^2}{m_{\psi}^2}\right)+20)+3M\Delta F\right)+3m_{\psi}^2\Delta L\right)\\ 
&M_8=-2E_l m_{\psi} M^2(2E_l-M)\left(E_lM(3F_2+3\log\left(\frac{u^2}{m_{\psi}^2}\right)-1)+12m_{\psi}^2\Delta F-3M^2\Delta F\right)
\end{align}
with,
\begin{align}
F_2=F_1\left((M^2-2k\cdot l);m_{\psi},m_{\psi}
\right),\quad \Delta F=F_1-F_2,
\end{align}
\begin{align}
A=4F_1+6\log\left(\frac{u^2}{m_{\psi}^2}\right)+7,\quad B=2F_1+2\log\left(\frac{u^2}{m_{\psi}^2}\right)+3
\end{align}
\begin{align}
{\rm Logsq}=\left[\log\left(\frac{\sqrt{M^4-4m_{\psi}^2M^2}+2m_{\psi}^2-M^2}{2m_{\psi}^2}\right)\right]^2
\end{align}
\begin{align}
{\rm Logsq2}=\left[\log \left(\frac{\sqrt{(M^2-2k\cdot l)(M^2-4m_{\psi}^2-2k\cdot l)}+2k\cdot l+2m_{\psi}^2-M^2}{2m_{\psi}^2}\right)\right]^2
\end{align}
\begin{align}
\Delta L=\rm Logsq-Logsq2
\end{align}
Using the same definitions, the quantities $M_9$ to $M_{16}$ are given by
\begin{align}
&M_{9}=-6m_{\psi}M\bigg(-4E_l^2M+4E_lm_{\psi}^2(2F_1-3F_2-2)\\ \nonumber
&\hspace{1cm}+2E_lM^2(-\Delta F+1)-M\Delta F(M^2-4m_{\psi}^2)\bigg)\\
&M_{10}=-6m_{\psi}E_l\left(-2E_lM^2\Delta F+8E_lm_{\psi}^2(F_1+2)+M(M^2-4m_{\psi}^2)\Delta F\right)\\
&M_{11}=-6E_lM^2(2E_l-M)\bigg(E_lM(3F_2+6\log\left(\frac{u^2}{m_{\psi}^2}\right)+8)+(3M^2-12m_{\psi}^2)\Delta F\bigg)\\
&M_{12}=-9E_lM^2(2E_l-M)\left(2E_lM(F_2+\log\left(\frac{u^2}{m_{\psi}^2}\right)+3)+M^2\Delta F+m_{\psi}^2\Delta L\right)\\
&M_{13}=-6E_l^2M^3(2E_l-M)\\
&M_{14}=-2E_lm_{\psi} M^2(2E_l-M)\left(2E_lM(3F_2+3\log\left(\frac{u^2}{m_{\psi}^2}\right)+5)-12m_{\psi}^2\Delta F+3M^2\Delta F\right)\\
&M_{15}=3E_lM^2(2E_l-M)\left(M\left(E_l(6F_2+6\log\left(\frac{u^2}{m_{\psi}^2}\right)+22)+3M\Delta F\right)+3m_{\psi}^2\Delta L\right)\\
&M_{16}=2E_l m_{\psi} M^2(2E_l-M)\left(E_lM(3F_2+3\log\left(\frac{u^2}{m_{\psi}^2}\right)-1)+12m_{\psi}^2\Delta F-3M^2\Delta F\right)
\end{align}
As argued in \cite{Barman:2023ymn}, in the rest frame of the inflaton, terms containing factors of the form $k^{n_{1}}e_{m_{1}n_{1}}(\boldsymbol{l},\chi)$ vanish identically. Consequently, the contributions from $\mathcal{M}_{\varphi(k)\to\bar{\psi}(p,s)\psi(q,s^{\prime})h_{m_{1}n_{1}}(l,\chi)}^{\text{(one-loop)},(i)}\big\vert_{\overline{\text{MS}}}$ with $(i=1,2,4)$ are zero. Only the terms with $(i=3,5)$ remain nonvanishing. Therefore, when evaluating the squared amplitude, it suffices to consider
\begin{align}
\nonumber
\vert\mathcal{M}_{\varphi(k)\to\bar{\psi}(p,s)\psi(q,s^{\prime})h_{ab}(l,\chi)}^{\text{(tree+one-loop)}}\vert^{2}&=\big\vert~\mathcal{M}_{\varphi(k)\to\bar{\psi}(p,s)\psi(q,s^{\prime})h_{ab}(l,\chi)}^{\text{(tree)}}\\
\nonumber
&+\mathcal{M}_{\varphi(k)\to\bar{\psi}(p,s)\psi(q,s^{\prime})h_{ab}(l,\chi)}^{\text{(one-loop)},(3)}\vert_{\overline{\text{MS}}}\\
&+\mathcal{M}_{\varphi(k)\to\bar{\psi}(p,s)\psi(q,s^{\prime})h_{ab}(l,\chi)}^{\text{(one-loop)},(5)}\vert_{\overline{\text{MS}}}~\big\vert^{2}
\end{align}This implies that the graviton is emitted solely from the external fermion legs. In other words, for the three-body decay process, graviton emission does not interfere with the loop integration. As a consequence, after applying dimensional regularization and the minimal subtraction scheme, the energy scale associated with the graviton, $E_l$, does not enter the terms arising from the loop corrections. For this reason, we assume that the inflaton two-body decay with one-loop corrections (corresponding to Fig.~\ref{TorsionOneLoopTwoBody}) and the three-body decay with one-loop corrections (corresponding to Fig.~\ref{TorsionOneLoopThreeBodyGWs}) share the same renormalization scale $u$, with $m_\psi \leq u \leq M$. Finally, after applying the various trace identities presented in Appendix~\ref{IdentityToFermiLoopIntegral}, we obtain the result given in Eq. \eqref{SquareAmpliThreeBodyTreeLoop}.

\printbibliography
\end{document}